\documentclass[%
 reprint,
superscriptaddress,
groupedaddress,
nofootinbib,
 amsmath,amssymb,
 aps,
floatfix,
]{revtex4-2}

\usepackage{graphicx}
\usepackage{dcolumn}
\usepackage{bm}
\usepackage{hyperref}
\usepackage{mathtools,mathrsfs}%

\usepackage{xspace}
\usepackage{xcolor}
\usepackage{booktabs}
\usepackage{makecell}
\usepackage{multirow}
\usepackage{threeparttable}
\usepackage[caption=false]{subfig}

\newcommand{\GeV}{\,\text{GeV}\xspace}
\newcommand{\TeV}{\,\text{TeV}\xspace}
\newcommand{\invfb}{\,\ensuremath{\mathrm{fb^{-1}}}\xspace}
\newcommand{\ttbar}{\ensuremath{t\overline{t}}\xspace}
\newcommand{\hbb}{\ensuremath{H\rightarrow bb}\xspace}
\newcommand{\xbb}{\ensuremath{X\rightarrow bb}\xspace}
\newcommand{\bbar}{\ensuremath{b\overline{b}}\xspace}
\newcommand{\pt}{\ensuremath{p_{\text{T}}}\xspace}
\newcommand{\kt}{\ensuremath{k_{\text{T}}}\xspace}
\newcommand{\HT}{\ensuremath{H_{\text{T}}}\xspace}

\newcommand{\ggF}{\ensuremath{gg\text{F}}\xspace}

\newcommand{\MGvATNLO}{\texttt{MG5\_aMC@NLO}\xspace}
\newcommand{\PYTHIA} {{\texttt{Pythia}}\xspace}
\newcommand{\VINCIA} {{\texttt{Vincia}}\xspace}
\newcommand{\POWHEG} {{\texttt{Powheg}}\xspace}
\newcommand{\HERWIG} {{\texttt{Herwig}}\xspace}
\newcommand{\DELPHES} {{\texttt{Delphes}}\xspace}
\newcommand{\JetClassII} {{\text{JetClass-II}}\xspace}
\newcommand{\Sophon} {\texttt{Sophon}\xspace}
\newcommand{\SophonAKFour} {\texttt{SophonAK4}\xspace}

\begin{document}

\preprint{APS/123-QED}

\title{\boldmath Potential of di-Higgs observation via a calibratable jet-free $HH\to 4b$ framework}

\author{Tianyi Yang}
\email{tianyi.y@cern.ch}
\affiliation{School of Physics and State Key Laboratory of Nuclear Physics and Technology, Peking University, 100871 Beijing, China}

\author{Congqiao Li}
\email{congqiao.li@cern.ch}
\thanks{Corresponding author}
\affiliation{School of Physics and State Key Laboratory of Nuclear Physics and Technology, Peking University, 100871 Beijing, China}

\date{\today}

\begin{abstract}
We present a calibratable, jet-free framework that enhances the search significance of the flagship LHC channel $HH \to 4b$ by more than a factor of five compared to existing approaches. The method employs a mass-decorrelated discriminant to identify $h_1 h_2 \to 4b$ with variable $h_{1,2}$ masses and a simultaneous estimator of $(m_{h_1},\,m_{h_2})$, both derived from multiclass classification on all-particle inputs. The $HH$ signal response can be calibrated using $ZZ \to 4b$. Using a highly realistic simulation framework validated through multiple tests, we demonstrate the method's robustness and identify two prerequisites essential for achieving this level of sensitivity. Results indicate that with LHC Run 2 and 3 data, observation-level sensitivity to $HH$ appears within reach, enabling constraints on $\kappa_\lambda$ comparable to HL-LHC projections and offering an accelerated path to precision measurements of the Higgs trilinear coupling.
\end{abstract}

\maketitle

\let\oldaddcontentsline\addcontentsline
\renewcommand{\addcontentsline}[3]{}

\textbf{\textit{Introduction.---}}
The observation of the Higgs boson by ATLAS and CMS in 2012~\cite{ATLAS:2012yve,CMS:2012qbp} completed the Standard Model (SM), yet left the shape of the Brout--Englert--Higgs potential largely untested. Expanding the scalar potential about its minimum through $V(\phi) \to V(v + h)$ yields 
\begin{equation}
    V = V_0 + \frac{1}{2}m_h^2 h^2 + \frac{m_h^2}{2v^2}vh^3 + \frac{1}{4}\frac{m_h^2}{2v^2}h^4,
\end{equation}
where the cubic term encodes the trilinear self-coupling $\lambda_{HHH} = m_h^2/2v^2$. This relation makes di-Higgs production, $pp\!\to\!HH$, a unique probe of the Higgs potential and the mechanism of electroweak symmetry breaking.
However, the SM predicts extremely small di-Higgs cross sections, with $31.1^{+2.1}_{-7.2}\,\text{fb}$ for gluon-gluon fusion (\ggF) and $1.73\pm0.04\,\text{fb}$ for vector-boson fusion (VBF) at $\sqrt{s}=13\TeV$~\cite{Dawson:1998py,Borowka:2016ehy,Baglio:2018lrj,deFlorian:2013jea, Shao:2013bz,deFlorian:2015moa,Grazzini:2018bsd,Baglio:2020ini,Dreyer:2018qbw}, making it an exceptionally challenging yet crucial process to explore for the HL-LHC era.

With the largest branching ratio, $HH \to b\overline{b}b\overline{b}$ (abbreviated $4b$) is the LHC's flagship channel for di-Higgs studies and has been extensively investigated in the literature~\cite{Dolan:2012rv,FerreiradeLima:2014qkf,Wardrope:2014kya,Behr:2015oqq,Amacker:2020bmn,Chiang:2024pho,Li:2024qfq,Wu:2025jza}. Recent searches are able to constrain the deviation of the self-coupling from the SM prediction, $\kappa_\lambda = \lambda_{HHH}/\lambda_{HHH}^{\text{SM}}$, to the few-unit level: with LHC Run~2 data ($\sim$140\invfb), ATLAS and CMS obtain 95\%~CL upper limits of $5.4$ and $3.9\,\sigma_{HH}^{\rm SM}$, yielding $\kappa_\lambda \in [-3.9,\,11.3]$ and $[-2.3,\,9.4]$~\cite{ATLAS:2023qzf,CMS:2022dwd}. These results arise from ``resolved analyses'' which target four resolved $b$ jets. Both experiments also pursue ``boosted analyses'', reconstructing $H\to b\overline{b}$ within large-radius jets~\cite{ATLAS:2024lsk,CMS:2023yay}. Remarkably, CMS boosted \ggF analysis achieves a comparable constraint, $\kappa_\lambda \in [-9.9,\,16.9]$ (expected $[-5.1,\,12.2]$)~\cite{CMS:2023yay}, despite relying on a topology less naturally suited to constrain $\kappa_\lambda$. The performance is driven largely by modern deep learning-based $H\to \bbar$ jet taggers based on jet constituents~\cite{CMS-DP-2020-002,Qu:2019gqs}, which dramatically suppresses multijet backgrounds~\cite{CMS:2023yay}. This raises an important question: can the superior background-rejection power of the boosted channel be extended across the broader resolved phase space?

In this \textit{Letter}, we report a major advance in di-Higgs sensitivity by deploying modern deep-learning techniques to the full trigger-accepted regime through a \textit{jet-free} $HH\to 4b$ framework. Using a realistic simulation, we input all pileup-mitigated reconstructed particles into a transformer network to distinguish generic $h_1 h_2 \to 4b$ signals from multijet backgrounds without relying on jet reconstruction. The flat mass prior embedded in the signal generation allows for calibration of the $HH$ response using $ZZ \to 4b$ events. For a projected 450\invfb dataset (Run~2+3), the method achieves an expected $2.8\,\sigma$ sensitivity under realistic systematic uncertainties---exceeding existing resolved analyses by a factor of $>$5 and approaching observation-level sensitivity. For reference, the latest ATLAS--CMS HL-LHC combination at $3\,\mathrm{ab^{-1}}$ reports $2.8\,\sigma$ in $HH \to 4b$~\cite{ATL-PHYS-PUB-2025-018}, compared with $1.0\,\sigma$ estimated from 2019~\cite{Dainese:2019rgk,ATLAS:2019mfr}. This work thus represents a major methodological advance, offering a fast-track path to precision measurements of the trilinear Higgs coupling years ahead of schedule.

\textbf{\textit{Simulation setup.---}}
Our method relies on low-level particle candidates, making an accurate description of their properties essential. We therefore develop a dedicated fast-simulation chain using \PYTHIA~8.3~\cite{Sjostrand:2014zea} for hadronization and \DELPHES~3.5~\cite{deFavereau:2013fsa} with a customized \JetClassII configuration~\cite{Li:2024htp}, which yields the full set of pileup-mitigated reconstructed particles. These are clustered into anti-\kt jets~\cite{Cacciari:2008gp} with $R = 0.4$ and 0.8 (jets and fatjets) following CMS conventions. As shown in \textit{Supplementary Material}~\cite{supp}, their realism relative to CMS ``particle-flow objects''~\cite{CMS-PRF-14-001,CMS:2020ebo} is validated using the \SophonAKFour and \Sophon transformer-based jet taggers: when trained on our particles, their performance matches that of leading CMS taggers~\cite{Zhao:2025rci,Li:2024htp}. These ``\Sophon taggers'' allow us to emulate the CMS resolved and boosted triggers and reproduce the corresponding analyses~\cite{CMS:2022dwd,CMS:2023yay}. Signal samples (SM \ggF, VBF, and $\kappa_\lambda$-varied signals) and backgrounds ($Z$+jets, $\ttbar$, single-top, diboson, single-Higgs) are generated with \MGvATNLO~2.9.18~\cite{Alwall:2014hca} or \POWHEG~2.0~\cite{Alioli:2008gx,Nason:2004rx,Frixione:2007vw} at leading order or next-to-leading order, showered with \PYTHIA, and normalized to the highest available cross-section accuracy~\cite{Czakon:2011xx,Kidonakis:2012rm,Gehrmann:2014fva,Bagnaschi:2011tu,Nason:2009ai,Hamilton:2012np,Luisoni:2013cuh,Grazzini:2018bsd,Dreyer:2018qbw,Alioli:2008gx,Nason:2004rx,Frixione:2007vw}. QCD multijet samples are produced by \PYTHIA with dedicated generator-level preselection to populate analysis-relevant regions and keep event generation tractable. All yields correspond to 450\invfb.

\begin{figure}[!ht]
\centering
\includegraphics[width=0.48\textwidth]{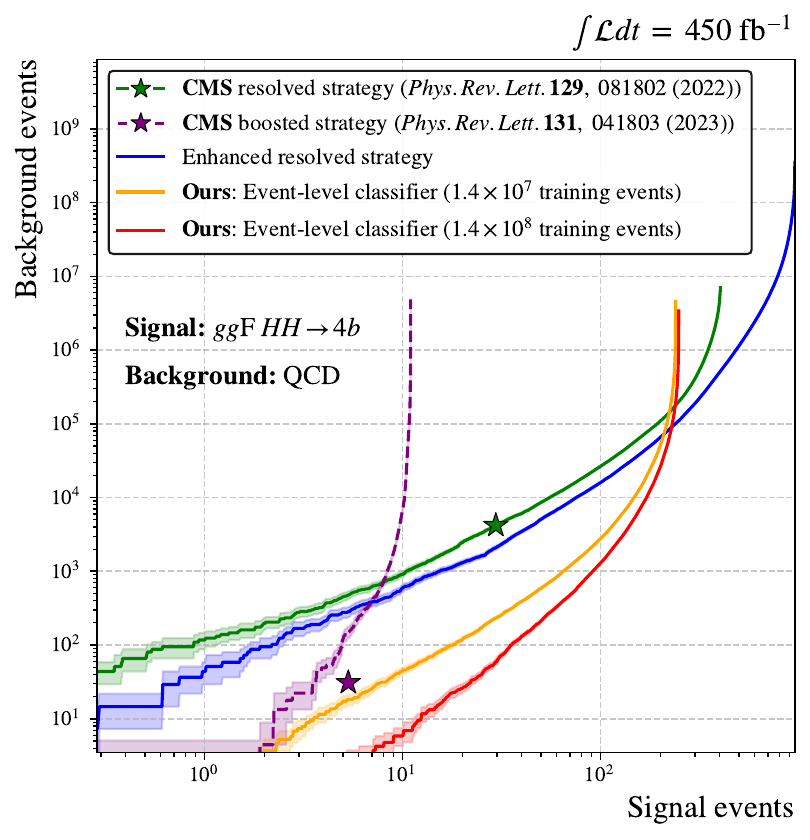}
\caption{
Receiver operating characteristic (ROC) curves for the $HH \to 4b$ signal versus the QCD background, scaled to the original event yields under 450\invfb. Shown are the reproductions of the CMS resolved-channel~\cite{CMS:2022dwd} (green) and boosted-channel~\cite{CMS:2023yay} (purple) strategies, enhanced resolved strategy probing the limit of jet-based approaches (blue), and our jet-free classifiers trained on the full $1.4\times 10^8$ events (red) or one-tenth of that sample (orange). True signal and background yields read from the signal-region figures in Refs.~\cite{CMS:2022dwd,CMS:2023yay} are marked by stars. Except for the boosted case, all strategies use resolved-channel triggers. Uncertainty bands indicate Poisson statistical limits. Our approach shows significant gains over established methods.
}
\label{fig:roc_comparison}
\end{figure}

Our first step is to faithfully reproduce the CMS resolved and boosted analyses to establish robust benchmarks. In the resolved channel~\cite{CMS:2022dwd}, the standard procedure selects four $b$‑tagged jets, forms two dijet pairs, and defines the signal region through their invariant masses. Replacing this two-step procedure, we deploy an end‑to‑end Particle Transformer (ParT)~\cite{Qu:2022mxj} operating on the four selected $b$-tagged jets, yielding the signal--background separation shown by the green curve in Fig.~\ref{fig:roc_comparison}. To probe the upper limit of jet-based resolved performance, we additionally include all jets in the event (including extra QCD radiation) and expose their \SophonAKFour flavor‑tagging scores as features; the resulting improvement is shown by the blue curve. The boosted channel~\cite{CMS:2023yay} selects \hbb candidates via the \Sophon \xbb discriminant and fatjet soft‑drop mass, corresponding to the purple dashed curve; although it triggers only about 1\% of the signal, it achieves substantially stronger background suppression. In both channels, the signal region event yields closely reproduce the experimental values (star markers), despite probing different regions of phase space, demonstrating the fidelity of our simulation.

\textbf{\textit{Methodology.---}}
Building on these foundations, we design a calibratable, jet-free strategy based on particle-level ParT networks ($\approx\!9\times10^6$ trainable parameters) that ingest all pileup‑mitigated reconstructed particles and perform event-level signal--background discrimination. The implementation is a 138‑class classifier. Of these, 136 signal classes correspond to the process $h_3\to h_1 h_2$ (produced within a two-Higgs-doublet model~\cite{Branco:2011iw} setup), discretizing $(m_{h_1},\,m_{h_2})$ over the range 40--200\GeV in 10\GeV increments, with $(m_{h_1},\,m_{h_2})$ and $(m_{h_2},\,m_{h_1})$ identified as the same class. The remaining two classes describe QCD multijet and \ttbar backgrounds. Training such a network is intrinsically challenging because the per‑event token multiplicity can reach $\mathcal{O}(300)$.

This tailored multiclass design provides two main advantages. (i) The aggregate of the 136 signal classes produces a flat prior in $(m_{h_1},\,m_{h_2})$, resulting in a mass‑decorrelated discriminant that is simultaneously sensitive to any $h_1 h_2\to 4b$ resonance structure within this plane, including the SM $ZZ$, $ZH$, and $HH$. (ii) By the Neyman--Pearson lemma~\cite{neyman1933ix}, the vector of per‑class signal scores serves as a discrete estimator of the underlying probability density ratio $p(m_{h_1},\,m_{h_2})$ for the hypothesis that an event arises from $h_1 h_2 \to 4b$ with masses $(m_{h_1},\,m_{h_2})$, from which a localized fit extracts $(\hat{m}_{h_1},\,\hat{m}_{h_2})$ as the best-fit peak position for each event. The classifier output thus provides both a powerful background-rejection discriminant, $D_{h_1 h_2\to 4b}$, and an event-level mass prediction point in the $(m_{h_1},\,m_{h_2})$ plane (up to exchange symmetry).

\begin{figure*}[htbp]
\centering
\includegraphics[width=\textwidth]{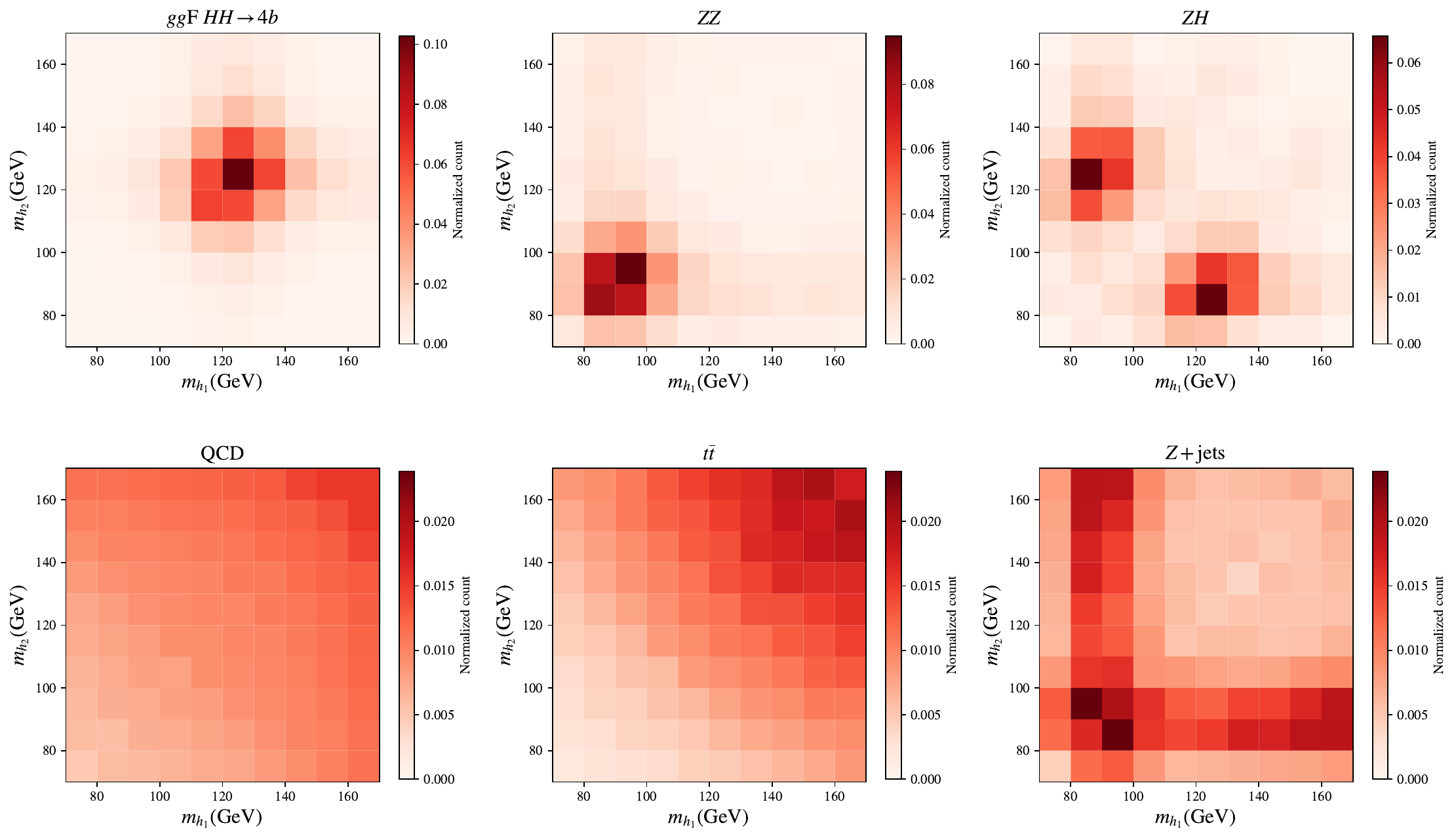}
\caption{Reconstructed $(m_{h_1},\,m_{h_2})$ distributions, normalized to unity, after applying a reference discriminant threshold $D_{h_1 h_2 \to 4b} > 0.9$. Six representative processes are shown: \ggF $HH\to 4b$, $ZZ$, $ZH$, QCD multijet, $\ttbar$, and $Z$+jets. The upper panels display signal-like processes, with reconstructed points forming the peak structure localized near $m_Z$ and $m_H$ along the two axes.}
\label{fig:mass_distributions}
\end{figure*}

Figure~\ref{fig:mass_distributions} shows the reconstructed $(m_{h_1},\,m_{h_2})$ distributions for six representative processes. Resonant signals such as \ggF $HH\to 4b$ and $ZZ\to 4b$ exhibit localized peaks near $(m_H,\, m_H)$ and $(m_Z,\, m_Z)$, respectively, while $ZH$ displays mixed structures. Events with a single $Z$ (or $H$) boson produce a double-ridge feature, while QCD and \ttbar backgrounds are broadly distributed without strong localized features.

The signal region is thus defined by a tight selection on $D_{h_1 h_2\to 4b}$ together with a mass selection requiring $(m_{h_1},\,m_{h_2})$ near $(m_H,\,m_H)$. To enhance stability and performance, we train an ensemble of three such classifiers. The final discriminant $D_{h_1 h_2 \to 4b}$ and the mass estimate are obtained by aggregating the outputs of the three ensemble members. As a result, the performance of the ensemble, shown by the red curve in Fig.~\ref{fig:roc_comparison}, \textit{markedly exceeds} that of existing resolved and boosted baselines: at high background rejection, the retained signal yield is larger by an order of magnitude or more, leading to a substantial gain in the search significance.

To identify the source of this improvement, we performed ablation studies and found that training statistics play a critical role. Our effective background sample contains $\sim$\,$6\times 10^7$ events, requiring the generation of $\sim$\,$2\times10^{11}$ unfiltered QCD events. Reducing the statistics by an order of magnitude degrades performance substantially (orange curve in Fig.~\ref{fig:roc_comparison}), highlighting the importance of large datasets for training high-performance discriminants on complex particle-level inputs.

\begin{figure*}[htbp]
\centering
\includegraphics[width=0.41\textwidth]{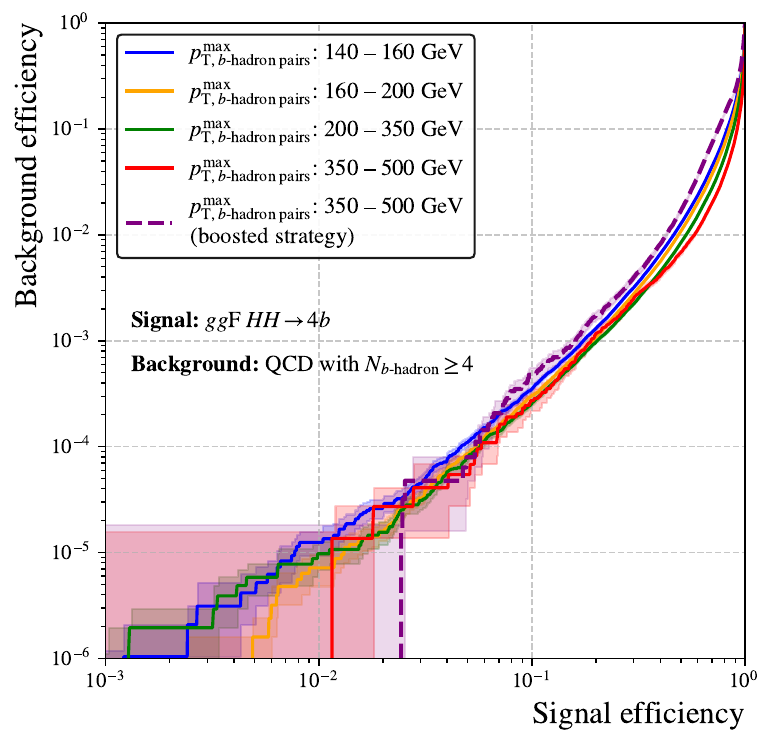}
\hspace{20pt}
\includegraphics[width=0.40\textwidth, trim={0 -6mm 0 0}, clip]{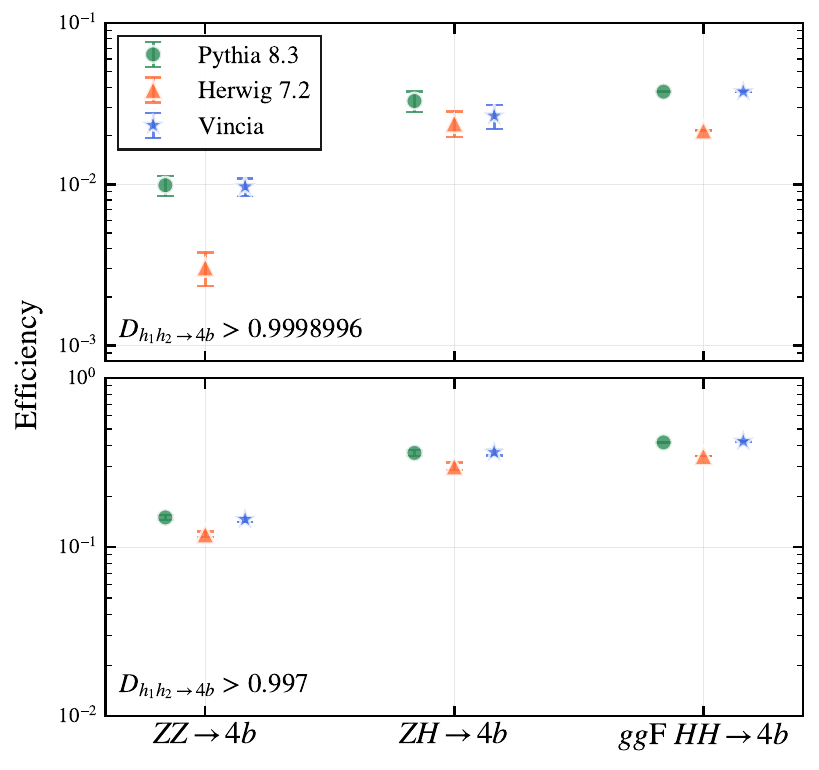}\label{fig:validation_b}
\caption{\textit{Left:} Validations of signal (\ggF $HH \to 4b$) versus background (QCD events with $\geq$4 $b$-hadrons) separation using ROC curves at different Lorentz boosts, characterized by maximum \pt among all $b$-hadron pair systems. Uncertainty bands show Poisson statistical limits. \textit{Right:} Comparison of three generators for hadronization modeling: \PYTHIA~8.3 (green), \HERWIG~7.2 (orange) and \VINCIA (blue) showing the $D_{h_1 h_2 \to 4b}$ selection efficiencies for \ggF $HH \to 4b$, $ZH \to 4b$ and $ZZ \to 4b$. The upper panel uses a threshold defining our signal region, while the lower panel applies a looser threshold.}
\label{fig:validation}
\end{figure*}

\textbf{\textit{Validation and calibration.---}}
It is crucial to verify that the significant sensitivity improvements arise from genuine methodological advances rather than artifacts of the simulation. We perform two validation studies. First, Fig.~\ref{fig:roc_comparison} shows that the signal and background yields obtained with our method (red curve) are approximately a rescaling of those from the CMS boosted strategy (purple curve). This motivates an explicit test that the classifier maintains comparable discriminating power for the SM $HH\to 4b$ signal versus background across different event Lorentz boosts.
Figure~\ref{fig:validation} (left) shows ROC curves obtained after partitioning events by a boost proxy, i.e., the maximum transverse momentum among all $b$-hadron pairs, using QCD events with exactly four $b$ hadrons as background. The classifier exhibits stable performance across all boost ranges. In the most Lorentz-boosted region, a direct comparison with the CMS boosted-jet-based strategy (purple curve in Fig.~\ref{fig:roc_comparison}) shows comparable signal--background discrimination. This confirms that the jet-free classifier retains the strong discriminating power characteristic of boosted approaches while extending it smoothly into the resolved regime.

Second, following standard LHC practice~\cite{CMS:2020poo,ATLAS:2025dkv}, we assess robustness to hadronization modeling by comparing \PYTHIA~8.3, \HERWIG~7.2~\cite{Bellm:2019zci}, and \VINCIA~\cite{Fischer:2016vfv}. As shown in Fig.~\ref{fig:validation} (right), differences increase at tighter working points, particularly for \HERWIG, consistent with previous observations\footnote{The reported \HERWIG--\PYTHIA differences are consistent with those seen experimentally (e.g., Ref.~\cite{ATL-PHYS-PUB-2023-020}) and are therefore expected. Moreover, recent taggers in ATLAS and CMS are routinely trained on \PYTHIA-showered samples and calibrated with data, with scale factors typically near unity~\cite{CMS:2025kje,CMS-DP-2025-010}.}. Yet, no significant efficiency degradation is observed in either alternative shower. This indicates that the classifier captures genuine features of the $h_1 h_2\to 4b$ final state. Further details are provided in \textit{Supplementary Material}~\cite{supp}.

Calibration of the classifier is central for experimental usage.
For the $HH$ signal, our strategy enables an experimentally robust ``calibrate--validate--measure'' workflow: derive simulation-to-data efficiency ratios (i.e., scale factors) using $ZZ\to 4b$ events near the $(m_Z,\,m_Z)$ peak, validate them on the $ZH$ structure, and apply them to $HH$. This mirrors the long-established use of $Z\to \bbar$ as a calibration ``candle'' for $H\to \bbar$ in boosted Higgs analyses~\cite{ATLAS:2023jdk,ATL-PHYS-PUB-2021-035,CMS:2025kje} and also aligns with an existing proposal employing $ZZ$ and $ZH$ to validate the $HH$ search strategy~\cite{CMS:2024tdk}. The consistency of efficiency ratios between generators in Fig.~\ref{fig:validation} (right) supports the transfer from $ZZ$ to $ZH$ and subsequently to $HH$. We further confirm that variations in the $Z$ boson spin (1 vs 0), which modify the angular distributions of the decay products, have a negligible impact on the classifier response for $ZZ\to 4b$.

For background calibration, several data-driven control regions can be defined. Estimating the QCD contribution is crucial, as it remains roughly an order of magnitude larger than all other backgrounds combined after the full event selection~\cite{supp}. Given its flat kinematic distributions, it can be naturally extracted from the mass sidebands in the $(m_{h_1},\,m_{h_2})$ plane. Notably, one can construct an additional control region using a separate classifier trained on $h_1 h_2 \to c\overline{c}c\overline{c}$ topology, which will be effectively free of $HH$ (and $ZH$) contamination and thus suitable for testing the background estimation strategy. A detailed realization of these strategies is deferred to their implementation within the LHC analyses.

\textbf{\textit{Results.---}}
We evaluate the discovery potential of our method by optimizing the $D_{h_1 h_2\to 4b}$ threshold and selecting events near $(m_H,\,m_H)$ in the $(m_{h_1},\,m_{h_2})$ plane. The expected significance is computed using the Asimov approximation with signal and background yields ($s,\,b$) and the background uncertainty $\sigma_b$~\cite{Cowan:2010js}. The 95\% CL constraint on $\kappa_\lambda$ is obtained via a scan using a profile likelihood ratio-based test statistic~\cite{CMS-NOTE-2011-005}, which incorporates $s$ (as a function of $\kappa_\lambda$), $b$, and their uncertainties $\sigma_s$ and $\sigma_b$.
Based on a fit-driven uncertainty estimate, we consider four benchmark scenarios with relative background (signal) uncertainties of $0\,(0)\%$, $5\,(10)\%$, $10\,(20)\%$, and $20\,(30)\%$. The resulting significances and 95\% CL intervals are summarized in Table~\ref{tab:results}. Notably, under a realistic uncertainty scenario, our method achieves a $2.8\,\sigma$ sensitivity, approaching the discovery threshold and representing a factor of $>$5 improvement over current jet-based resolved analyses. The ensemble technique provides a 10\% improvement beyond the individual model performances, which range from 2.2--2.6. With 450\invfb from a single experiment, the sensitivity already reaches the ATLAS--CMS HL-LHC projections obtained with traditional methods~\cite{ATL-PHYS-PUB-2025-018,Dainese:2019rgk,ATLAS:2019mfr}.

\begin{table}[htbp]
\centering
\caption{Summary of the expected SM di-Higgs statistical significance and 95\% CL constraints on $\kappa_\lambda$ under different signal and background uncertainty scenarios, for an integrated luminosity of 450\invfb.}
\vspace{2pt}
\begin{tabular}{ccc}
\toprule
Background (signal) & SM significance & 95\% CL $\kappa_\lambda$ \\
uncertainties & ($\sigma$) & range \\
\midrule
0\% (0\%) & 3.3 & $[-0.59,\,8.5]$ \\
5\% (10\%) & 3.1 & $[-0.81,\,8.7]$ \\
10\% (20\%) & 2.8 & $[-1.3,\,9.3]$ \\
20\% (30\%) & 2.1 & $[-2.1,\,10.0]$ \\
\bottomrule
\end{tabular}
\label{tab:results}
\end{table}

\textbf{\textit{Conclusion.---}}
In summary, we demonstrate in this \textit{Letter} that the $HH \to 4b$ channel offers far greater discovery potential than previously recognized. By exploiting the full particle content of collision events, our approach achieves unprecedented background rejection and delivers an expected $2.8\,\sigma$ sensitivity with 450\invfb under realistic systematics---already competitive with HL-LHC projections based on conventional strategies~\cite{ATL-PHYS-PUB-2025-018,Dainese:2019rgk,ATLAS:2019mfr}. The method constrains $\kappa_\lambda \in [-0.81, 8.7]$ at 95\% CL, thereby enabling precision probes of the Higgs self-coupling years ahead of schedule.

Although the direct use of low-level particle information may appear aggressive for immediate experimental deployment, a central contribution of this work is the development of a complete and calibratable analysis strategy, constructed with reference to experimentally-grounded strategies at the LHC. In brief, the framework features a mass-decorrelated $h_1 h_2 \to 4b$ ``event tagger'', enabling $HH$ signal calibration through the $ZZ$ process and intermediate validation through $ZH$, and an estimator for $(m_{h_1},\,m_{h_2})$ to extract peak-like structures in the mass spectrum. We deliver several validation studies to confirm the reliability of the method.

Our results, particularly those shown in Fig.~\ref{fig:roc_comparison}, demonstrate that the significant performance gain relies critically on (i) going beyond jet-based analyses and (ii) the use of sufficiently large training datasets supported by dedicated deep-learning engineering. Although several alternative proposals have explored sensitivity improvements in $HH \to 4b$~\cite{Amacker:2020bmn,Chiang:2024pho,Li:2024qfq,Wu:2025jza}, to our knowledge none have uncovered the level of gain realized here, which we attribute to these two key ingredients.

Looking forward, this approach opens new opportunities for precision Higgs physics and beyond. The demonstrated sensitivity suggests that observation-level reach for SM di-Higgs production may already be attainable with existing LHC datasets, with further gains expected from upgraded $HH\to 4b$ resolved trigger strategies introduced since Run~3~\cite{ATLAS:2025nyf,CMS-DP-2023-050,CMS-DP-2025-009}, as well as from continued scaling of the training sample size and model capacity, which we find have not yet reached saturation. We show that the achievable precision on $\kappa_\lambda$ reaches levels previously thought accessible only at the HL-LHC, thereby opening the possibility of studies that were long considered out of reach, such as tri-Higgs production~\cite{Abouabid:2024gms}. More broadly, our particle-level strategy provides a scalable template for next-generation searches, in which full-event analyses, empowered by advances in deep learning---larger models, richer training samples, and greater computing power---promise to boost a variety of signal searches and access previously unexplored regions of parameter space.

\textbf{\textit{Data availability.---}}
The code and trained models used in this study are publicly available~\cite{code}. The simulated datasets are not hosted due to size constraints but are available from the authors upon request.

\textbf{\textit{Acknowledgement.---}}
The authors appreciate helpful discussions with Raghav Kansal, Qiang Li, Huilin Qu, Liaoshan Shi and Lei Zhang. CL thanks colleagues in the ATLAS and CMS Collaborations for valuable discussions and feedback during the Higgs Pairs Workshop at Isola d'Elba. CL and TY thank Yipin Wang and Chen Zhou for valuable discussions on the robust evaluation of ROC performance.
This research is supported in part by the computational resource operated at the Institute of High Energy Physics of the Chinese Academy of Sciences.

\bibliography{main.bib}

\clearpage
\onecolumngrid
\appendix
\makeatletter
\let\addcontentsline\oldaddcontentsline
\setcounter{equation}{0}
\setcounter{figure}{0}
\setcounter{table}{0}
\setcounter{page}{1}
\renewcommand{\theequation}{S\arabic{equation}}
\renewcommand{\thefigure}{S\arabic{figure}}
\renewcommand{\thetable}{S\Roman{table}}

\begin{center}
{\large \textbf{\boldmath Supplementary Material for ``Potential of di-Higgs observation via a calibratable jet-free $HH\to 4b$ framework''}}\\
\vspace{15pt}
Tianyi\, Yang\, and\, Congqiao\, Li\\
{\small \textit{School of Physics and State Key Laboratory of Nuclear Physics and Technology, Peking University, 100871 Beijing, China}}\\
{\small (Dated: \today)}
\end{center}

\tableofcontents

\section{Supplementary details on simulation framework and \texttt{Sophon} tagger performance}\label{app:sophon-tagger-perf}

As discussed in the main text, this work relies on using a high-fidelity fast simulation to generate pileup-mitigated particles that closely match those in real experiments. The key validation step is to reproduce jet and fatjet (or large-$R$ jet) taggers on this simulation and verify that their performance aligns with the leading CMS taggers. This section summarizes the simulation setup and the corresponding tagger reproduction results.

\subsection{Simulation framework}

The fast-simulation workflow is built on a modified \DELPHES configuration derived from the default CMS card~\cite{deFavereau:2013fsa}, incorporating two key changes. First, we include a module that applies transverse $d_0$ and longitudinal $d_z$ impact-parameter smearing as functions of the jet \pt and $\eta$. The smearing prescriptions follow those of a Run~1 CMS configuration. Since $d_0$ and $d_z$ serve as inputs to the jet-tagging neural network, critical for heavy-flavor identification, it is essential that their resolutions reflect realistic tracking performance. Second, pileup is modeled assuming an average of 50 additional interactions per event, and a pileup per-particle identification (PUPPI) algorithm~\cite{Bertolini:2014bba} is employed to remove pileup contributions, using parameters retuned for our custom fast-simulation environment. This procedure results in a set of pileup-mitigated reconstructed particles that closely reproduces CMS conditions; we refer to this setup as the \JetClassII\ \DELPHES configuration~\cite{Li:2024htp}. Jets with radii $R=0.4$ and $R=0.8$ are then clustered from these particles using the anti-\kt algorithm~\cite{Cacciari:2008gp}.

\subsection{The \texttt{Sophon} tagger for \texorpdfstring{$bb$}{bb} identification}

For the large-$R$ jet tagger \texttt{Sophon}, we build on the methodology of Ref.~\cite{Li:2024htp}, ``Signature-Oriented Pre-training for Heavy-resonance ObservatioN''. In this framework, a jet model is pretrained as a large-scale multiclass classifier on the comprehensive \JetClassII dataset~\cite{jetclass2}. The pretraining task distinguishes among 188 large-$R$ jet signatures in a mass- and \pt-decorrelated manner, covering a wide range of resonant topologies (2-, 3-, and 4-prong, including the $X\to bb$ channel) as well as QCD jet final states. This methodology provides a prototype for next-generation boosted-jet taggers in CMS. Notably, the recently developed Global Particle Transformer (\texttt{GloParT}) follows a similar multiclass and mass/\pt-decorrelated design (see Refs.~\cite{CMS-PAS-HIG-23-012,CMS-PAS-JME-25-001,CMS:2025rqr,CMS-PAS-HIG-24-010}).

The model architecture is built on the Particle Transformer (ParT)~\cite{Qu:2022mxj} backbone with 6 particle-attention blocks and 2 class-attention blocks, featuring a 128-dimensional embedding space and 8 attention heads. The fully connected multilayer perceptron (MLP) is expanded to two layers with dimensions (512, 188), where the 128-dimensional neuron values before the final MLP serve as latent features for transfer learning applications. The full model contains approximately $2.3\times 10^6$ parameters.

Jet classes for \texttt{Sophon} are defined as follows. Resonant jets are initiated from generic spin-0 resonances $X$ decaying to various final states: two-prong signatures include diparton decays ($bb$, $cc$, $ss$, $qq$, $bc$, $cs$, etc.) and dilepton decays ($ee$, $\mu\mu$, $\tau_h\tau_e$, etc.); three- and four-prong signatures arise from cascade decays $X \to Y^{(*)}Y^{(*)}$ with subsequent diparton/dilepton decays of the secondary resonances. QCD background jets are subdivided into 27 classes based on the number and flavors of quarks within the jet.

\begin{figure}[ht!]
\centering
\includegraphics[width=0.44\textwidth]{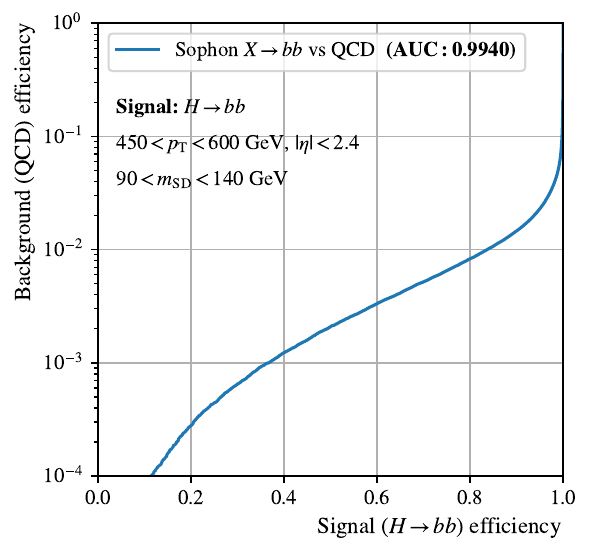}
\hfill
\caption{Performance of the \texttt{Sophon} model $H\to bb$ versus QCD jets, shown as ROC curves with AUC values. The performance is directly comparable to established CMS large-$R$ \xbb jet taggers under identical phase-space selection~\cite{CMS:2025kje}.}
\label{fig:sophon_performance}
\end{figure}

For $bb$ tagging, the optimal discriminant for distinguishing $X \to bb$ signals from QCD backgrounds is constructed using the likelihood-ratio properties of multiclass classifiers,
\begin{equation}\label{eq:sophon-xbb}
\text{discr}(X \to bb \text{ vs QCD}) = \frac{g_{X \to bb}}{g_{X \to bb} + \sum_{l=1}^{27} g_{\text{QCD}_l}},
\end{equation}
where $g_{X \to bb}$ corresponds to the $X \to bb$ output score and $g_{\text{QCD}_l}$ represents the scores of the 27 QCD classes.
The $bb$-tagging performance is evaluated using the receiver operating characteristic (ROC) curve in Fig.~\ref{fig:sophon_performance}, with Standard Model (SM) \hbb jets as signal and QCD jets as background, following CMS conventions~\cite{CMS:2025kje}. Table~\ref{tab:bkgrej_sophon} lists the background rejection ($1/\epsilon_B$) at signal efficiencies $\epsilon_S=60\%$ and $40\%$. For comparison, we include the CMS taggers \texttt{DeepDoubleX}, \texttt{ParticleNet-MD} and \texttt{GloParT} using results from Refs.~\cite{CMS:2025kje,CMS-PAS-HIG-24-010}. The performance of \texttt{Sophon} lies between \texttt{DeepDoubleX} and \texttt{ParticleNet-MD} and is closer to the latter, though it does not reach the discrimination power of \texttt{GloParT}, likely reflecting residual differences between fast and full detector simulations. These results, however, demonstrate that the $bb$-tagging performance reproduced on our dedicated fast-simulation dataset is close to that of CMS's leading taggers.

\begin{table}[!ht]
\centering
\caption{Comparison of the \Sophon model with the established CMS taggers including \texttt{DeepDoubleX}, \texttt{ParticleNet-MD} and \texttt{GloParT} on $bb$-tagging performance. Results for \texttt{DeepDoubleX} and \texttt{ParticleNet-MD} are taken from Ref.~\cite{CMS:2025kje}, Fig.~5. The \texttt{GloParT} rejection rate corresponds to the latest version used in recent CMS $HH\to 4b$ search~\cite{CMS-PAS-HIG-24-010} and is derived by applying the \texttt{GloParT}-to-\texttt{ParticleNet-MD} rejection ratio from Ref.~\cite{CMS-PAS-HIG-24-010}, Fig.~5, to the \texttt{ParticleNet-MD} values.}
\vspace{5pt}
\label{tab:bkgrej_sophon}
\begin{tabular}{lcccc}
\toprule
  & \DELPHES\ (\JetClassII) simulation & \multicolumn{3}{c}{CMS simulation~\cite{CMS:2025kje,CMS-PAS-HIG-24-010}} \\
\cmidrule(lr){2-2} \cmidrule(lr){3-5}
  & ~~\Sophon~~ & ~~\texttt{DeepDoubleX}~~ & ~~\texttt{ParticleNet-MD}~~ & ~~\texttt{GloParT} \\ 
\midrule
$X\to bb$ vs.~QCD, $\epsilon_{\rm s} = 60\%$ & 300 & 200 & 370 & 520 \\ 
$X\to bb$ vs.~QCD, $\epsilon_{\rm s} = 40\%$ & 810 & 580 & 970 & 1520 \\ 
\bottomrule
\end{tabular}
\end{table}

\subsection{The \texttt{SophonAK4} tagger for \texorpdfstring{$b$}{b}-jet identification}

For small-$R$ jet tagging, we introduce \SophonAKFour, which adopts the same design philosophy as the \texttt{Sophon} tagger but is applied to anti-\kt jets with $R=0.4$. This enables a realistic treatment of $b$-tagging even in fast-simulation studies: rather than relying on a binary $b$-tag flag, we obtain continuous discriminant scores for $b$ and other flavor scenarios. These scores can be used to emulate the CMS trigger strategy, define custom analysis $b$-tagging working points in offline selections, or even serve as inputs to event-level neural-network training.

\SophonAKFour is trained on a broad set of small-$R$ jets drawn from a dataset generated with the same fast-simulation workflow as the \JetClassII configuration: \MGvATNLO for matrix-element generation, \PYTHIA~8.3 for parton showering and hadronization, and \DELPHES~3.5 for detector simulation using the \JetClassII data card. The tagger employs a streamlined ParT-based architecture optimized for the small-$R$ jet tagging. It incorporates 6 particle-attention blocks and 2 class-attention blocks with a 64-dimensional embedding and 8 attention heads, for a total model size of roughly $5.5\times 10^5$ trainable parameters.

The \SophonAKFour model used in this work corresponds to the second iteration (v2), building on the initial version introduced in Ref.~\cite{Zhao:2025rci}. The primary upgrade in v2 is an improved labeling scheme that unifies ``hadron flavor'' and ``parton flavor'' information, derived using a procedure analogous to the CMS ghost-matching algorithm. For $b$-tagging (where the hadron flavor is identified as $b$), this scheme enables a direct comparison with the CMS small-$R$ jet taggers.

The ghost-matching algorithm associates heavy-flavor hadrons ($b$ and $c$), leptons, and partons to reconstructed jets by reclustering the jet constituents together with ``ghost'' particles, i.e., truth-level particles rescaled by a small factor $\epsilon$ so as not to affect jet kinematics while enabling flavor association. For each jet, the algorithm assigns:
\begin{itemize}
\item \textit{Hadron flavor}: Based on the presence of $b$ hadrons (\texttt{hadronFlavor} = 5), $c$ hadrons (\texttt{hadronFlavor} = 4), or neither (\texttt{hadronFlavor} = 0);
\item \textit{Parton flavor}: Based on the leading partons or lepton pairs matched to the jet, categorized into 26 classes covering quarks, gluons, and leptonic signatures.
\end{itemize}

The complete \texttt{SophonAK4} classification scheme includes 32 classes organized by hadron flavor:
\begin{itemize}
\item The B class (\texttt{hadronFlavor} = 5, classes 0--4) includes single $b$-quark ($b$, $\overline{b}$), $b$-quark pairs ($b\overline{b}$), other parton combinations, and invalid signatures.
\item The C class (\texttt{hadronFlavor} = 4, classes 5--9) follows an analogous structure with $c$-quark signatures ($c$, $\overline{c}$, $c\overline{c}$), other combinations, and invalid cases.
\item The L class (\texttt{hadronFlavor} = 0, classes 10--31) includes a broader range of signatures: fake heavy flavor misidentifications (class 10), single light quarks and gluons ($s$, $\overline{s}$, $d$, $\overline{d}$, $u$, $\overline{u}$, $g$; classes 11--17), single leptons ($e^{\pm}$, $\mu^{\pm}$, $\tau_h^{\pm}$; classes 18--23), light quark pairs ($s\overline{s}$, $d\overline{d}$, $u\overline{u}$, $gg$; classes 24--27), lepton pairs ($e^+e^-$, $\mu^+\mu^-$, $\tau_h^+\tau_h^-$; classes 28--30), and final state radiation with other signatures (class 31).
\end{itemize}

The discriminant construction then exploits this detailed classification to achieve optimal separation between different flavor categories:
\begin{equation}
\text{B-score} = \sum_{i=0}^{4} p_i, \quad \text{C-score} = \sum_{i=5}^{9} p_i, \quad
\text{L-score} = \sum_{i=10}^{17} p_i + \sum_{i=24}^{27} p_i + p_{31}
\end{equation}
where $p_i$ represents the output probability for class $i$.

\begin{figure}[ht!]
\centering
\includegraphics[width=0.44\textwidth]{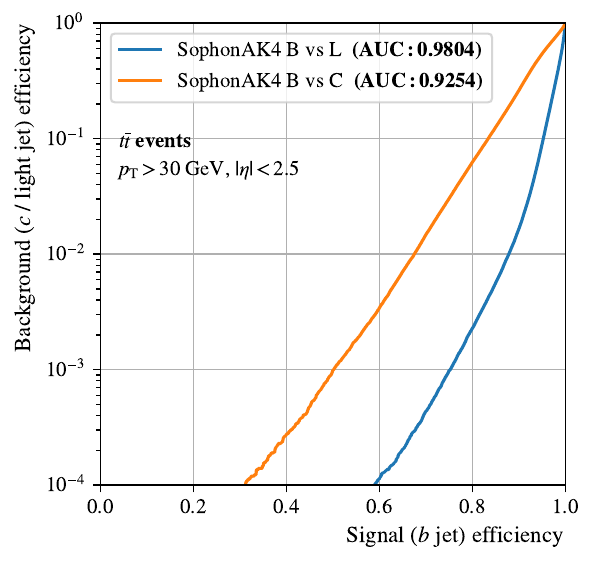}
\includegraphics[width=0.44\textwidth]{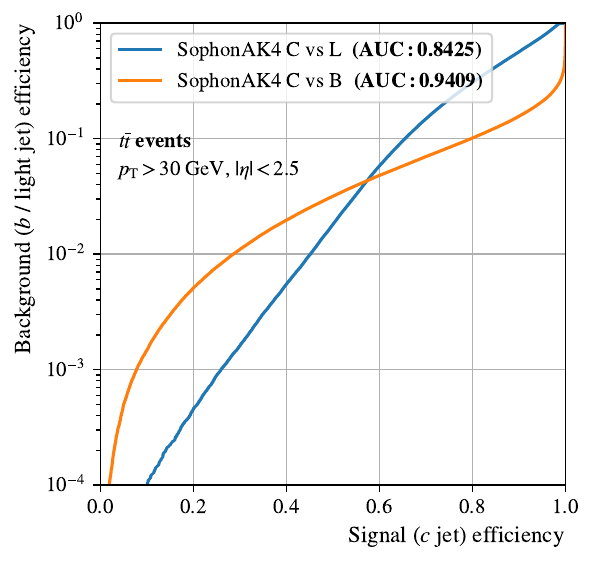}
\caption{Performance of the \texttt{SophonAK4} model (v2) for (left) $b$ versus light/charm jet tagging and (right) $c$ versus light/$b$ jet tagging, shown as ROC curves with AUC values. Signal and background jets originate from \ttbar events with jets satisfying $\pt > 30$~GeV and $|\eta| < 2.5$. The performance is directly comparable to established CMS flavor taggers under identical phase-space selections.}
\label{fig:sophon_ak4_performance}
\end{figure}

Figure~\ref{fig:sophon_ak4_performance} presents the ROC curves for \texttt{SophonAK4} performance on $b$ versus light/charm jet tagging and $c$ versus light/$b$ jet tagging, evaluated on SM \ttbar events with jets satisfying $\pt > 30$~GeV and $|\eta| < 2.5$, allowing for a direct comparison with the performance of CMS taggers, including \texttt{DeepJet}, \texttt{ParticleNet}, and \texttt{UParT} (from Ref.~\cite{CMS-DP-2024-066}, Figs.~1 and 3). The background rejections ($1/\epsilon_B$) derived from the ROC curves are summarized in Table~\ref{tab:bkgrej_sophonak4}. Interestingly, the model demonstrates similar performance with established CMS taggers: for $b$ versus light jet discrimination, \texttt{SophonAK4} achieves performance between \texttt{ParticleNet} and \texttt{UParT} levels, while for charm tagging applications, it reaches performance comparable to or exceeding current CMS standards.

\begin{table}[!ht]
\centering
\caption{Comparison of the \SophonAKFour model (v2) with the established CMS taggers, \texttt{DeepJet}, \texttt{ParticleNet}, and \texttt{UParT}, regarding $b$ vs.~light jet tagging and $c$ vs.~light/$b$ jet tagging performance. The background rejection rate ($1/\epsilon_B$) is shown for $b$ vs.~light jet tagging at fixed signal efficiencies of $\epsilon_b = 80\%$ and $60\%$, and $c$ vs.~light/$b$ jet at fixed signal efficiencies of $\epsilon_c = 60\%$ and $40\%$. The CMS results are extracted from the ROC curves in Ref.~\cite{CMS-DP-2024-066}, Figs.~1 and 3.}
\vspace{5pt}
\label{tab:bkgrej_sophonak4}
\begin{tabular}{lcccc}
\toprule
  & \DELPHES simulation & \multicolumn{3}{c}{CMS simulation~\cite{CMS-DP-2024-066}} \\
\cmidrule(lr){2-2} \cmidrule(lr){3-5}
  & ~~\SophonAKFour~~ & ~~\texttt{DeepJet}~~ & ~~\texttt{ParticleNet}~~ & ~~\texttt{UParT} \\ 
\midrule
$b$ vs.~light jet, $\epsilon_{\rm b} = 80\%$ & 440 & 220 & 350 & 390 \\ 
$b$ vs.~light jet, $\epsilon_{\rm b} = 60\%$ & 8800 & 4100 & 6700 & 10000 \\ 
\midrule[1pt]
$c$ vs.~light jet, $\epsilon_{\rm c} = 60\%$ & 17 & 9.8 & 16.4 & 16.0 \\ 
$c$ vs.~light jet, $\epsilon_{\rm c} = 40\%$ & 180 & 78 & 190 & 210 \\ 
\midrule
$c$ vs.~$b$ jet, $\epsilon_{\rm c} = 60\%$ & 20.8 & 16.4 & 22.1 & 24.0 \\ 
$c$ vs.~$b$ jet, $\epsilon_{\rm c} = 40\%$ & 51 & 36 & 49 & 55 \\ 
\bottomrule
\end{tabular}
\end{table}

The validation studies confirm that both \texttt{Sophon} and \texttt{SophonAK4} provide realistic tagging performance consistent with experimental benchmarks. In particular, these results confirm that particle-level inputs provide a sufficiently realistic description for our purposes, thus supporting the credibility of the full-event analysis strategy employed in this work.

\section{Supplementary details on reproduction of established $HH\to 4b$ strategies}\label{app:CMS-perf}

This section summarizes the implementation of the CMS resolved and boosted strategies, as illustrated in Fig.~\ref{fig:roc_comparison}. Demonstrating that our framework accurately reproduces the existing CMS measurements is an essential prerequisite for establishing the validity of the jet-free analysis strategy developed in this work. We begin by outlining the setup of the Monte Carlo samples and trigger selections, followed by two subsections describing the reproduction of the resolved and boosted analyses.

\subsection{Physics samples and trigger selections}\label{app:samples_and_triggers}

Our work employs a comprehensive suite of SM background processes and di-Higgs signal samples to replicate the conditions of the corresponding ATLAS and CMS measurements. Table~\ref{tab:sample_preparation} summarizes the full set of simulated datasets. Below, we describe the relevant generator configurations, special settings, trigger definitions, and event statistics for various samples.

\begin{table}[!ht]
\centering
\caption{Monte Carlo samples and trigger selection statistics for resolved channel analysis. Event counts show the progression through resolved channel trigger requirements: 4-jet kinematic selection ($\pt > [75,\,60,\,45,\,40]$\GeV, $\HT > 330$\GeV) and loose $b$-tagging requirement (3 $b$-tagged jets) for an integrated luminosity of 450\invfb. See the text for details.}
\label{tab:sample_preparation}
\vspace{5pt}
\begin{threeparttable}
\begin{tabular}{p{0.20\linewidth}p{0.22\linewidth}p{0.16\linewidth}p{0.12\linewidth}p{0.12\linewidth}p{0.12\linewidth}}
\toprule
\textbf{Process} & \textbf{Generator} & \textbf{\boldmath Total events \phantom{xx} (450\invfb)}~\tnote{$*$} & \textbf{After trigger ``4j''} & \textbf{After trigger ``3b''} & \textbf{Trigger efficiency} \\
\midrule
\multicolumn{6}{c}{\textbf{Signal processes}} \\
\midrule
\ggF $HH$ ($\kappa_\lambda = \kappa_t = 1$) & \POWHEG & 4,734 & 1,107 & 959 & 20.2\% \\
VBF $HH$ & \MGvATNLO LO & 263 & 37 & 27 & 10.1\% \\
\midrule
\multicolumn{6}{c}{\boldmath \textbf{QCD and $Z+$jets backgrounds}} \\
\midrule
QCD multijet & \PYTHIA & $2.03 \times 10^{12}$~\tnote{$\dagger$} & $6.39 \times 10^8$ & $3.50 \times 10^8$ & 0.017\% \\
$Z \to qq$ & \MGvATNLO LO+012j & $3.00 \times 10^8$~\tnote{$\ddagger$} & $2.85 \times 10^6$ & $1.46 \times 10^6$ & 0.49\% \\
\midrule
\multicolumn{6}{c}{\textbf{Top quark backgrounds}} \\
\midrule
\ttbar & \MGvATNLO LO+01j & $3.74 \times 10^8$ & $3.69 \times 10^7$ & $1.66 \times 10^7$ & 4.4\% \\
Single top ($t$-, $s$-channel) & \MGvATNLO LO+01j & $8.33 \times 10^7$ & $1.90 \times 10^6$ & $8.06 \times 10^5$ & 0.97\% \\
$tW$ & \MGvATNLO LO+01j & $2.93 \times 10^7$ & $1.17 \times 10^6$ & $5.63 \times 10^5$ & 1.9\% \\
$t\overline{t}W$ & \MGvATNLO LO & $3.35 \times 10^5$ & $4.99 \times 10^4$ & $2.33 \times 10^4$ & 7.0\% \\
$t\overline{t}Z$ & \MGvATNLO LO & $3.87 \times 10^5$ & $7.83 \times 10^4$ & $4.19 \times 10^4$ & 10.8\% \\
$t\overline{t}H$ & \POWHEG & $2.28 \times 10^5$ & $7.38 \times 10^4$ & $4.64 \times 10^4$ & 20.3\% \\
\midrule
\multicolumn{6}{c}{\textbf{Diboson backgrounds}} \\
\midrule
$WW$ & \MGvATNLO LO+01j & $5.36 \times 10^7$ & $5.04 \times 10^4$ & $3.52 \times 10^4$ & 0.066\% \\
$WZ$ & \MGvATNLO LO+01j & $2.10 \times 10^7$ & $7.88 \times 10^4$ & $4.26 \times 10^4$ & 0.20\% \\
$ZZ$ & \MGvATNLO LO+01j & $7.61 \times 10^6$ & $2.94 \times 10^4$ & $1.52 \times 10^4$ & 0.20\% \\
\midrule
\multicolumn{6}{c}{\textbf{Single Higgs backgrounds}} \\
\midrule
\ggF $H$ & \POWHEG & $2.19 \times 10^7$ & $5.99 \times 10^4$ & $2.34 \times 10^4$ & 0.11\% \\
VBF $H$ & \POWHEG & $1.70 \times 10^6$ & $1.38 \times 10^4$ & $5.13 \times 10^3$ & 0.30\% \\
$W^+H$ & \POWHEG & $3.78 \times 10^5$ & $6.80 \times 10^3$ & $3.22 \times 10^3$ & 0.85\% \\
$W^-H$ & \POWHEG & $2.40 \times 10^5$ & $4.17 \times 10^3$ & $2.03 \times 10^3$ & 0.85\% \\
$ZH$ & \POWHEG & $3.42 \times 10^5$ & $6.71 \times 10^3$ & $3.44 \times 10^3$ & 1.0\% \\
\bottomrule
\end{tabular}%
\begin{tablenotes}
\footnotesize
\item[$*$] To estimate the event yields, all samples, except QCD and $Z$+jets, use cross sections computed at the highest available perturbative order from the corresponding theoretical literature.
\item[$\dagger$] This value corresponds to the \PYTHIA cross section for events generated with \texttt{pTHat} required to exceed 75\GeV.
\item[$\ddagger$] This value corresponds to the \MGvATNLO cross section for $Z\to qq$ samples generated with up to two additional partons and the dedicated parton-level phase-space selection on the leading three partons, as described in the text. The quoted numbers include the effects of matrix-element and parton-shower merging and represent the post-matching cross sections.
\end{tablenotes}

\end{threeparttable}
\end{table}

\vspace{10pt}
\textbf{\textit{Generation details.}}---
The signal samples comprise $HH$ production via \ggF and VBF, generated using \POWHEG~2.0 at next-to-leading-order (NLO) accuracy and \MGvATNLO~2.9.18~\cite{Alwall:2014hca} at leading-order (LO) accuracy, respectively. In addition to the SM samples, grids with varied values of $\kappa_\lambda$ (including 0 and 5) are produced to enable characterization of non-SM scenarios and to support the profiling and limit-setting procedures. The simulated backgrounds include inclusive QCD multijet production, $Z$+jets production, top-quark processes (\ttbar, single top, $tW$, $t\overline{t}W$, $t\overline{t}Z$, $t\overline{t}H$), electroweak diboson production ($WW$, $WZ$, $ZZ$), single-Higgs production across all major channels (\ggF, VBF, $VH$). Among these, the single-Higgs processes (including $t\overline{t}H$) are generated with \POWHEG~2.0 at NLO accuracy in QCD~\cite{Alioli:2008gx,Nason:2004rx,Frixione:2007vw}. The QCD multijet samples are produced with \PYTHIA, while the remaining background processes are generated with \MGvATNLO~2.9.18~\cite{Alwall:2014hca} at LO accuracy, employing the \texttt{NNPDF}~3.1 NNLO PDF set~\cite{Ball:2017nwa}. Up to one or two partons are included at the matrix-element level before matching and merging with the parton shower.

To maintain computational efficiency for the most demanding backgrounds, notably QCD multijet and $Z$+jets, dedicated generator-level phase-space preselections are applied to enrich regions relevant for the analysis. For QCD multijet production, we generate $2\to 2$ hard-scattering events with the \texttt{pTHat} scale required to exceed 75\GeV. At the \PYTHIA level, we further require at least four generator-level jets (clustered from stable final-state particles) satisfying $\pt > (70,\,50,\,35,\,35)$\GeV, a scalar \pt sum of all stable particles exceeding 350\GeV, and a generator-level \HT (scalar sum of jet \pt for jets with $\pt > 20$\GeV) above 240\GeV. For $Z$+jets, the \MGvATNLO configuration includes up to two partons at matrix-element level, with the $Z$ boson decaying to two quarks in the hard process. We require the \pt of the three leading matrix-element partons to exceed $(70,\,50,\,35)$\GeV, and their scalar \pt sum to exceed 180\GeV. All preselection thresholds are optimized to ensure that no phase-space regions that can pass the online trigger requirements are inadvertently removed.

\vspace{10pt}
\textbf{\textit{Cross sections.}}---
With the exception of QCD and $Z$+jets, the cross sections are taken from the highest-order perturbative calculations available, as reported in Refs.~\cite{Czakon:2011xx,Kidonakis:2012rm,Gehrmann:2014fva,Bagnaschi:2011tu,Nason:2009ai,Hamilton:2012np,Luisoni:2013cuh,Grazzini:2018bsd,Dreyer:2018qbw,Alioli:2008gx,Nason:2004rx,Frixione:2007vw}. For the QCD and $Z$+jets multijet samples, the cross sections are estimated using the leading-order predictions from \PYTHIA and \MGvATNLO, respectively, with dedicated phase-space selections described above.

Using these cross-section inputs, the expected event yields for an integrated luminosity of 450\invfb are summarized in Table~\ref{tab:sample_preparation}. Because our study targets sensitivity in regimes where backgrounds are suppressed to exceptionally low levels, substantial Monte Carlo statistics are required, particularly for the dominant backgrounds: QCD multijet, $Z$+jets, and \ttbar. In our simulation, the per-event weights needed to scale these samples to a reference integrated luminosity of 100\invfb are approximately 1.6, 0.3, and 0.5, respectively. This implies an effective raw QCD sample size of order $10^{11}$--$10^{12}$ events, which is an intrinsically challenging requirement. For all remaining processes, the event weights are kept below 1.0.

\vspace{10pt}
\textbf{\textit{Trigger requirements.}}---
Trigger emulation follows the standard CMS menu requirements. The resolved channel requires $\geq 4$ jets (at reconstruction-level) with \pt thresholds of $(75,\,60,\,45,\,40)$\GeV, a scalar $\HT>330$ GeV, and $\geq 3$ jets passing a loose \SophonAKFour $b$ tag. The loose working point corresponds to a light-jet misidentification rate of $\epsilon_{B}=10\%$, implemented via a \SophonAKFour B-score threshold of 0.0243. This configuration reproduces the conventional ``4j3b'' trigger strategy used by CMS in Run 2~\cite{CMS:2022dwd}. We verify that the resulting trigger rate is consistent with expectations: for 450\invfb, the simulated QCD multijet yield is $3.50\times10^{8}$ events, corresponding to a trigger rate of 18~Hz at an instantaneous luminosity of $2\times10^{34}\,\mathrm{cm^{-2}s^{-1}}$. The top-right point of the blue curve in Fig.~\ref{fig:roc_comparison} indicates the background level after this trigger selection.

The trigger efficiencies vary substantially across processes, as shown in Table~\ref{tab:sample_preparation}. The highest efficiencies are observed for Higgs-pair production (\ggF $HH$: 20.2\%, VBF $HH$: 10.1\%) and for top-associated processes ($t\overline{t}H$: 20.3\%, $t\overline{t}Z$: 10.8\%). Standard top backgrounds show moderate efficiency (\ttbar: 4.4\%), while QCD multijet and $Z$+jets processes have significantly lower rates (0.017--0.49\%).

For the boosted channel, we adopt the standard CMS requirements of scalar $\HT>800$\GeV for all large-$R$ jets with \pt exceeding 200\GeV, and a leading large-$R$ jet with trimmed mass $m_{\mathrm{trim}}>50$\GeV, consistent with the CMS Run 2 $HH\to 4b$ analysis that we aim to reproduce~\cite{CMS:2023yay}.

\subsection{Resolved-channel strategy}

\subsubsection{A simple cut-based baseline}

We begin with a simple cut-based analysis that reconstructs Higgs-boson pairs from four $b$-tagged jets. For events passing the resolved trigger path introduced above, we select four offline $b$-tagged jets with $\pt > 30\GeV$, $|\eta| < 2.5$, and satisfying the \SophonAKFour medium working point (corresponding to a light-jet misidentification rate of $\epsilon_B = 1\%$ and a B-score threshold of 0.187).

The four jets with the highest \SophonAKFour B-scores are then used for Higgs reconstruction. Given four $b$-jets, $j_1,\,j_2,\,j_3,\,j_4$, the three possible pairings to form two Higgs candidates are $\{\{j_1,j_2\},\{j_3,j_4\}\}$, $\{\{j_1,j_3\},\{j_2,j_4\}\}$, $\{\{j_1,j_4\},\{j_2,j_3\}\}$.
Following the strategy commonly employed in ATLAS and CMS $HH\to 4b$ analyses~\cite{ATLAS:2023qzf,CMS:2022dwd}, we compute for each pairing the diagonal distance
\begin{equation}
d_{HH} = \frac{|m_{H_1} - \alpha\,m_{H_2}|}{\sqrt{1+\alpha^2}},
\end{equation}
where $\alpha = 1.04$ accounts for the slight asymmetry in the reconstructed Higgs masses. If the smallest and second-smallest diagonal distances differ by less than 30~GeV, the pairing with the larger leading-Higgs $\pt$ in the di-Higgs rest frame is chosen.

The signal region is defined as a circular window centered at $(m_{H_1},\,m_{H_2}) = (110,\,105)\GeV$ with radius 25\GeV. In contrast to CMS, which centers the region near the true Higgs mass, we do not apply $b$-jet energy-regression corrections. Consequently, the reconstructed masses exhibit a systematic downward shift, concentrating signal events at lower invariant masses. The signal region is therefore given by
\begin{equation}
\sqrt{(m_{H_1} - 110)^2 + (m_{H_2} - 105)^2} < 25\GeV.
\end{equation}

This cut-based selection yields 197 expected signal events and 302,778 background events at an integrated luminosity of 450\invfb. This operating point lies above the green and blue curves in Fig.~\ref{fig:roc_comparison}, serving as a baseline that reflects the most elementary resolved-analysis strategy.

\subsubsection{End-to-end machine-learning baselines}

We benchmark the performance attainable with the resolved strategy using an end-to-end advanced machine-learning approach. The goal is to assess the discrimination power between the \ggF $HH$ signal and the dominant QCD multijet background. These results correspond to the green and blue curves in Fig.~\ref{fig:roc_comparison}.

\vspace{10pt}
\textbf{\textit{\boldmath Four leading $b$-tagged jets.}}---
As in standard resolved analyses, once four $b$-tagged jets are selected, machine-learning algorithms can be employed to pair the jets (e.g., as explored in Refs.~\cite{FerreiradeLima:2014qkf,Wardrope:2014kya,Behr:2015oqq,Amacker:2020bmn}), correct their kinematics to improve the reconstructed di-Higgs mass resolution $(m_{H_1},\,m_{H_2})$, and define an optimal signal region on the mass plane. In place of this multi-step procedure, we train an end-to-end ParT model that takes the full kinematic information of the four selected $b$-tagged jets and directly performs the $HH$ versus QCD discrimination. The inputs are jet four-vectors $(p_x,\,p_y,\,p_z,\,E)$. The ParT backbone, designed for particle-level representations, efficiently constructs nodewise features (e.g., $\log(\pt)$ and $\log(E)$) and pairwise features (e.g., angular distances $\Delta R_{ij}$ and dijet invariant masses $m_{ij}$)~\cite{Qu:2022mxj}, thereby providing a direct path to the discrimination limit achievable with current machine-learning techniques.

For training, we adapt the ParT architecture to the four-token input using a compact network with embedding dimensions (64, 256, 64), pair-embedding dimensions (32, 32, 32), 8 particle-attention blocks, and 2 class-attention blocks with latent dimension 64, and 4 attention heads. The network has 2 output nodes for binary classification and is trained using the \texttt{Weaver} framework with the \texttt{Ranger} optimizer (learning rate $10^{-3}$, weight decay $0.01$), 20 epochs, batch size 512, and automatic mixed precision. Each epoch processes $5.12\times 10^5$ training samples and $1.28\times 10^5$ validation samples.

The resulting \ggF $HH$ versus QCD separation is shown by the green curve in Fig.~\ref{fig:roc_comparison} in the main text. We validate this benchmark by comparing with the signal--background separation reported by CMS. In Ref.~\cite{CMS:2022dwd}, \textit{Supplementary Material}, Fig.~5 (top-right panel), which corresponds to the most sensitive phase-space region, the deep neural network score is used to extract signal events. Selecting events in the highest 8 bins of that distribution yields the maximal expected sensitivity, corresponding to 6.7 signal events and 950 background events. After scaling to 450\invfb, this point (shown as the red marker in Fig.~\ref{fig:roc_comparison}) closely matches our reproduction of the resolved analysis. We note that a clearer visual comparison is provided in our public presentations; see Ref.~\cite{yang_2025_15459612}.

\vspace{10pt}
\textbf{\textit{All jets with full flavor information.}}---
We further examine the limit attainable within the resolved strategy. Recent studies have proposed including more than four jets in object-assignment networks (e.g., Refs.~\cite{Chiang:2024pho,Li:2024qfq}), since jets originating from Higgs decays may fail $\pt$ or $b$-tag requirements, while additional jets from QCD radiation may still carry discriminating information. We therefore extend the input to include up to ten jets per event satisfying $\pt > 30\GeV$ and $|\eta| < 2.5$, and augment the kinematic features with the B-, C-, and L-score from \SophonAKFour. These scores are entered as their logarithms to improve numerical stability. This configuration allows the network to access all potential Higgs-decay candidates, radiation patterns that differ between signal and background, and the full flavor information for each jet. The same architecture and training configuration described above is used. We note that this strategy is conceptually similar to the recent CMS search~\cite{CMS:2025dsh}, which leverages full jet-based kinematics and flavor information within an advanced architecture. While that analysis discretizes tagging scores to ensure calibratability, our goal here is to assess the attainable sensitivity limit, and we therefore provide the raw scores directly as inputs.

The resulting improvement is shown by the blue curve in Fig.~\ref{fig:roc_comparison}. Because no four-$b$-tag preselection is applied, the ROC curve begins at the inclusive yields after the resolved trigger. Relative to the green curve, the extended-jet network improves background rejection by a factor of roughly 2--3 at fixed signal efficiency, though the improvement appears modest on a logarithmic scale. This behavior indicates that \textit{jet-only resolved strategies have intrinsic limitations in background suppression}, and that substantially larger gains, as described later, require inputs that extend beyond jets alone.

\subsection{Boosted-channel strategy}

The boosted channel targets high-$\pt$ Higgs bosons whose decay products are collimated into large-$R$ jets, providing sensitivity to di-Higgs production at high invariant masses $m_{HH}$ where the natural boost leads to overlapping decay products. The analysis uses anti-\kt jets with radius parameter $R = 0.8$, groomed with the soft-drop algorithm ($\beta = 0$, $z_{\text{cut}} = 0.1$)~\cite{Larkoski:2014wba}. Building on the boosted-trigger selection discussed above, which imposes requirements on large-$R$ jet \HT and a loose mass threshold, the offline selection demands exactly two large-$R$ jets with $\pt > 200$\GeV, $|\eta| < 2.5$, and soft-drop mass $m_{\text{SD}} > 80$\GeV, which closely follows the boosted $HH \to 4b$ searches in ATLAS and CMS~\cite{CMS:2023yay,ATLAS:2024lsk}. The two highest-$\pt$ jets satisfying these criteria are used for di-Higgs reconstruction.

\subsubsection{A cut-based baseline for boosted channel}

To reproduce the boosted strategy, we adopt a simplified signal-region definition based on a circular mass window in the $(m_{\text{SD}}^{j_1},\,m_{\text{SD}}^{j_2})$ plane. The window is centered at $(125,\,115)$\GeV with radius 25\GeV:
\begin{equation}
\sqrt{(m_{\text{SD}}^{j_1} - 125)^2 + (m_{\text{SD}}^{j_2} - 115)^2} < 25\GeV.
\end{equation}
The asymmetric center reflects the expected difference in mass resolution between leading and subleading large-$R$ jets. The resulting \ggF\ $HH$ and QCD yields after the boosted-jet selection and mass window correspond to the starting point of the purple curve in Fig.~\ref{fig:roc_comparison}, indicating that only about 1\% of the events retained in the resolved analysis survive the boosted selection.

A simple discriminant is then constructed using the \Sophon model. For each jet, the \xbb versus QCD discriminant $D_{bb}$ from Eq.~(\ref{eq:sophon-xbb}) is evaluated, and the event-level score is defined as
\begin{equation}\label{eq:sophon-discr}
D_{\text{event}} = D_{bb}^{(j_1)} + D_{bb}^{(j_2)}.
\end{equation}
The purple curve in Fig.~\ref{fig:roc_comparison} shows the resulting signal--background separation, demonstrating a markedly stronger performance than in the resolved case: background rejection improves by three orders of magnitude while retaining roughly 90\% of the signal.

Importantly, the fidelity of this powerful separation performance can be validated against CMS measurements. In Ref.~\cite{CMS:2023yay}, Fig.~1, the reconstructed mass of $j_2$ is used for signal extraction. Selecting the central four bins (110--150\GeV) yields approximately 1.6 signal and a remarkably small 4.1 QCD multijet events at 138~fb$^{-1}$; after rescaling to 450~\invfb, this point is shown as the purple marker in Fig.~\ref{fig:roc_comparison}. CMS achieves about a factor of five stronger background rejection than our reproduction, which is understandable since their analysis employs an additional event-level multivariate approach that incorporates both large-$R$ jet kinematics and raw components of the $bb$-tagging discriminant, thereby enhancing separation power compared to our simpler cut-based baseline. We therefore conclude that, although the purple curve shows remarkably strong discrimination, \textit{this level of performance is well supported by existing experimental results}.

\section{Supplementary details on the jet-free \texorpdfstring{$HH\to 4b$}{HH->4b} approach}\label{app:training}

The main text presents a concise overview of the proposed methodology. The quantitative performance and robustness of the approach, however, rely on the detailed implementation, in particular on the design choices related to deep-learning training and data preparation. This appendix thus documents the training configuration in full technical detail. The description is intended to provide sufficient information for an independent reproduction and for potential deployment within the ATLAS or CMS experiments.

\subsection{Jet-free model development}

\subsubsection{Training samples and label definitions}

The model training strategy is guided by three key principles:
(1) the use of a very large training dataset, such that the background rejection capability in extreme phase-space regions scales with increasing statistics;
(2) the use of variable-mass $h_1 h_2 \to 4b$ processes as signal, enabling the model to generalize across resonant $4b$ final states, including $ZZ$, $ZH$, and $HH$ topologies;
and (3) a mass-based multiclass classification scheme that allows for an end-to-end estimation of the joint probability density in the $(m_{h_1},\,m_{h_2})$ plane for each event.
The concrete implementation details are described below.

\vspace{10pt}
\textbf{\textit{\boldmath Signal sample configuration.}}---
Signal events are generated using the \MGvATNLO implementation of a two-Higgs-doublet model via the process $pp \to h_3 \to h_1 h_2$.
The masses of the two Higgs bosons, $h_1$ and $h_2$, are sampled uniformly in 5\GeV increments over the range $[40,\,200]$\GeV.
The heavy scalar $h_3$ is required to satisfy the kinematic constraint $m_{h_3} > m_{h_1} + m_{h_2}$, and its mass distribution is chosen to provide broad coverage of the accessible phase space.

\begin{figure}[ht!]
\centering
\includegraphics[width=0.90\textwidth]{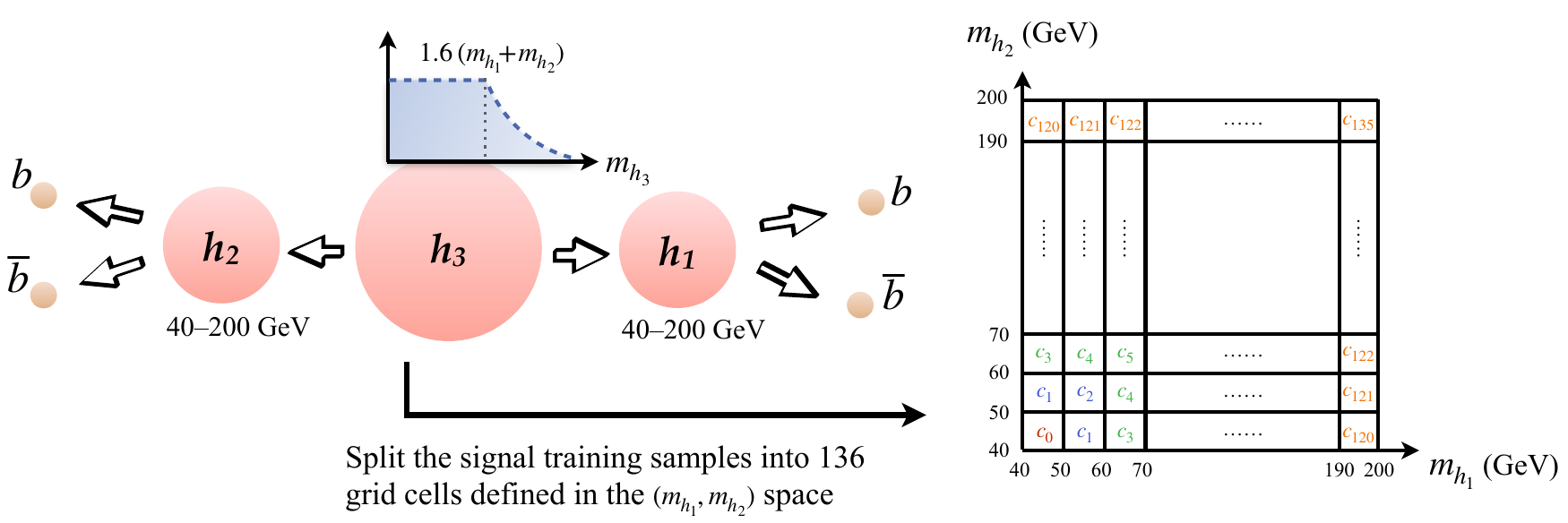}
\caption{A schematic illustration of the $h_1 h_2\to 4b$ signal sample design used in our method. The masses of $h_1$ and $h_2$ are uniformly sampled in the range of 40--200\GeV, whereas their invariant mass follows a predefined distribution. The signal events are classified into 136 grid cells in the $(m_{h_1},\,m_{h_2})$ parameter space, accounting for permutational symmetry.}
\label{fig:app_signal_illustration}
\end{figure}

In the SM and its extensions, the $m_{HH}$ spectrum typically exhibits a long-tailed behavior, with shape variations induced by the interference between triangle and box contributions for different values of $\kappa_\lambda$. Rather than modeling a specific coupling scenario, a flexible \textit{ad hoc} prescription is adopted for the sampling of $m_{h_3}$. For a given invariant mass sum $m_{h_1} + m_{h_2}$, the $h_3$ mass is generated according to a two-regime scheme:
\begin{itemize}
\item \textit{Uniform regime}: $m_{h_3} \in [m_{h_1} + m_{h_2},\, 1.6\,(m_{h_1} + m_{h_2})]$, sampled from a uniform distribution;
\item \textit{Exponential regime}: $m_{h_3} \in [1.6\,(m_{h_1} + m_{h_2}),\, +\infty)$, sampled from an exponential distribution with base $1/2$ and a characteristic scale of $0.4\,(m_{h_1} + m_{h_2})$.
\end{itemize}
The two regimes are matched continuously at $m_{h_3} = 1.6\,(m_{h_1} + m_{h_2})$ to ensure a continuous distribution. A schematic illustration of the signal sample design is shown in Fig.~\ref{fig:app_signal_illustration}.

To restrict the training to the phase space relevant for the analysis, we require events to pass at least one of two trigger selections and take their union. The first is the ``4j3b'' trigger used in the resolved analysis (see Appendix~\ref{app:samples_and_triggers}), requiring $\geq$3 jets to pass a loose $b$-tag working point with light-jet efficiency $\epsilon_B=10\%$. The second is the ``4j2b'' trigger, requiring $\geq$2 jets to pass a tight $b$-tag working point with $\epsilon_B=0.1\%$ (corresponding to a \SophonAKFour B-score threshold of 0.643). The latter is motivated by the updated ATLAS and CMS trigger menus, which employ tighter two-$b$-tag requirements~\cite{ATLAS:2025nyf,CMS-DP-2023-050,CMS-DP-2025-009}. Although the ``4j2b'' trigger is not explicitly studied in our work, taking the union of the two triggers both populates the training sample and better reflects a realistic experimental implementation.

In total, $5.0\times 10^8$ signal events are generated at the matrix-element level, of which $7.0\times 10^7$ (70\,M) events pass the trigger selection and are used for model training.

\vspace{10pt}
\textbf{\textit{\boldmath Background sample configuration.}}---
The background samples consist of QCD multijet and \ttbar processes, which constitute the dominant background contributions in the $HH\to 4b$ search. These samples are generated following the same procedures as those used for the inference samples described in Appendix~\ref{app:samples_and_triggers}, while remaining statistically orthogonal to the datasets used for model inference. The raw event counts for QCD and \ttbar samples are $2.0\times 10^{11}$ and $4.0\times 10^7$. These roughly match the actual event yields of the two processes according to the expected event yields presented in Table~\ref{tab:sample_preparation}. Similar to signal samples, for the QCD sample, only events that pass the union of ``4j3b'' or ``4j2b'' trigger paths are retained to construct the training dataset, resulting in $6.3\times 10^7$ (63\,M) events. To maintain a larger \ttbar training set, we impose only the ``3b'' or ``2b'' requirements on the \ttbar sample, i.e., applying the $b$-tagging selections without additional constraints on the leading four jet \pt thresholds or on \HT. This results in $1.2\times 10^7$ (12\,M) \ttbar events. 

In total, our training dataset comprises $1.44\times 10^8$ (144\,M) events, with comparable contributions from signal and background samples.

\vspace{10pt}
\textbf{\textit{\boldmath Label construction.}}---
We define 136 signal classes by discretizing the generator-level Higgs mass pair $(m_{h_1},\,m_{h_2})$ on a two-dimensional grid, as shown in Fig.~\ref{fig:app_signal_illustration}. Each class corresponds to a $10\GeV \times 10\GeV$ cell in the $(m_{h_1},\,m_{h_2})$ plane, with the exchange symmetry $(m_{h_1},\,m_{h_2}) \leftrightarrow (m_{h_2}, m_{h_1})$ explicitly taken into account. Exploiting this symmetry reduces the number of distinct signal classes from 256 to 136, while still allowing the full discrete probability density $p(m_{h_1},\,m_{h_2})$ on the original $16\times16$ grid to be reconstructed from the network outputs. The QCD multijet and \ttbar backgrounds are assigned to two additional classes, labeled 137 and 138.

It is worth emphasizing the motivation for adopting a multiclass formulation. A seemingly natural alternative would be to predict the event location in the $(m_{h_1},\,m_{h_2})$ plane directly in an end-to-end manner, as the experimental objective is to fill events into this plane after event selections. In existing ATLAS and CMS analyses, such a regression technique is typically used to reconstruct jet \pt or mass. However, an important statistical consideration is that regression models can only learn a central tendency of the underlying probability distribution for a given event (e.g., the mean or median) rather than the full probability density. This limitation is particularly severe in the present case, where the $(m_{h_1},\,m_{h_2})$ distribution is intrinsically multimodal and may exhibit more complex structures (e.g., see Fig.~\ref{fig:fitting_validation}). As a result, a regression-based approach becomes inadequate. By contrast, multiclass classification enables a direct estimation of the probability density through the multiclass cross-entropy loss. We therefore adopt an end-to-end probability estimation, followed by a rule-based grid assignment using a peak-finding procedure applied to the reconstructed distribution (as detailed in Appendix~\ref{app:fit-strategy}).

From another perspective, a multiclass workflow allows the model to learn optimal decision boundaries between any pair of classes. In the binary case, the Neyman--Pearson lemma~\cite{neyman1933ix} implies that the optimal classifier is given by the likelihood ratio of the corresponding probability densities. With a sufficiently expressive Transformer architecture, this optimality can be reached across all pairs of binary classification tasks, as demonstrated in Ref.~\cite{Li:2024htp}. Therefore, the model employed in this work simultaneously achieves optimal signal--background discrimination (equivalent to training a dedicated signal--background binary classifier) and an optimal estimate of the probability density across different signal hypotheses (equivalent to classifying among multiple signal components).

\subsubsection{Model architecture and input design}

We adopt ParT~\cite{Qu:2022mxj} as the baseline architecture, which treats particle-level features as Transformer tokens, learns their correlations, and outputs event-level classifications for signal and background. Our analysis further employs an ensemble of three independently trained ParT models. As will be seen, this ensemble strategy improves robustness and reduces statistical fluctuations in both signal--background discrimination and mass reconstruction.

\vspace{10pt}
\textbf{\textit{\boldmath Network configuration.}}---
We use an enlarged ParT model with a latent space twice the standard size, corresponding to approximately $9\times 10^6$ trainable parameters. The Transformer embedding dimensions are (256, 1024, 256), and the pairwise embedding dimensions are (64, 64, 64). The pairwise embedding size is not doubled (remaining at $d=64$) to control computational cost, which scales as $\mathcal{O}(N^2 d^2)$ and becomes significant for inputs with up to $N=300$ particles. The architecture consists of 8 particle-attention blocks and 2 class-attention blocks, each with latent dimension 256 and 16 attention heads. The fully connected layer has dimension 1024 with 10\% dropout regularization, followed by a 138-dimensional classification head with softmax activation.

\vspace{10pt}
\textbf{\textit{\boldmath Input variable configuration.}}---
The model processes full particle-flow information through synchronized input streams in the \JetClassII data format. We note that the same particle-level inputs are used to train the jet-based tagging models, \Sophon and \SophonAKFour. In total, 18 particle-level node-wise features are used:
\begin{itemize}
\item \textit{Kinematic variables}: $\ln \pt$, $\ln E$, $\ln(\pt^{\text{rel}})$, $\ln(E^{\text{rel}})$, $\Delta R$, $\Delta\eta$, $\Delta\phi$, $\eta$, $\phi$;
\item \textit{Track parameters}: $d_0$, $d_0^{\text{err}}$, $d_z$, $d_z^{\text{err}}$, with a hyperbolic tangent transformation applied to impact parameters;
\item \textit{Particle identification}: binary flags for charged hadrons, neutral hadrons, photons, electrons, and muons, together with a continuous charge variable.
\end{itemize}

As in the \Sophon and \SophonAKFour trainings, the same variable standardization is applied: logarithmic \pt and energy variables are centered at 1.7 and 2.0, respectively, with scale factors of 0.7; relative momentum variables are centered at $-4.7$ with the same scaling. Angular variables are normalized using a dedicated scheme, with $\Delta R$ centered at 0.2 and scaled by 4.0.

Following the standard ParT setup, the model also takes pairwise particle features as inputs: $\{\ln \kt,\, \ln z,\, \ln \Delta R,\, \ln m_{ij}\}$. Here, $\kt = p_{\text{T},\min}\,\Delta R$ is the \kt clustering metric, $z = p_{\text{T},\min}/(p_{\text{T},i}+p_{\text{T},j})$ encodes the momentum fraction, $\Delta R_{ij}=\sqrt{\smash[b]{(\Delta\eta_{ij})^2+(\Delta\phi_{ij})^2}}$ measures the angular separation, and $m_{ij}$ denotes the invariant mass of particle pairs $\{i,j\}$. These features carry useful Lorentz-invariance properties and play a key role in improving performance relative to plain Transformer architectures.

\subsubsection{Model training}

\vspace{10pt}
\textbf{\textit{\boldmath Training statistics.}}---
Model training uses the \texttt{Ranger} optimizer~\cite{liu2019variance}, a combination of \texttt{RAdam} and \texttt{Lookahead}~\cite{zhang2019lookahead}, with an initial learning rate of $2\times 10^{-3}$ and automatic mixed precision for improved computational efficiency. We employ a distributed data-parallel training scheme across four Nvidia A100 GPUs. The training is performed for 80 epochs with a batch size of 512. In each epoch, $10{,}000\times1024/4 = 2.56\times10^6$ samples are used for training and $2{,}500\times1024/4 = 6.4\times10^5$ samples for validation, where the factor of 4 corresponds to the number of GPUs. As a reference, the total training time for each model is about 92 hours. Three models are trained independently with identical architectures and hyperparameters but different random initializations to form the ensemble.

\vspace{10pt}
\textbf{\textit{\boldmath Studies on statistical balancing.}}---
We investigate whether sample reweighting is beneficial during training, as commonly done in high-energy physics to account for different physics processes or to decorrelate specific kinematic variables. After detailed studies, no event-level weights are applied. For the QCD and \ttbar backgrounds, the available training statistics closely match their expected event yields; thus, we do not prefer to apply additional sample reweighting. We also explore mass-decorrelation reweighting between signal and background samples and find that such procedures effectively reduce the usable training statistics and degrade model performance. Due to our dedicated design of the training samples, the distributions are already sufficiently smooth across all mass-related observables while maintaining a reasonable balance between signal and QCD background. Consequently, no event weights are applied to either the signal or background samples.

\begin{figure}[!ht]
\centering
\includegraphics[width=0.7\textwidth]{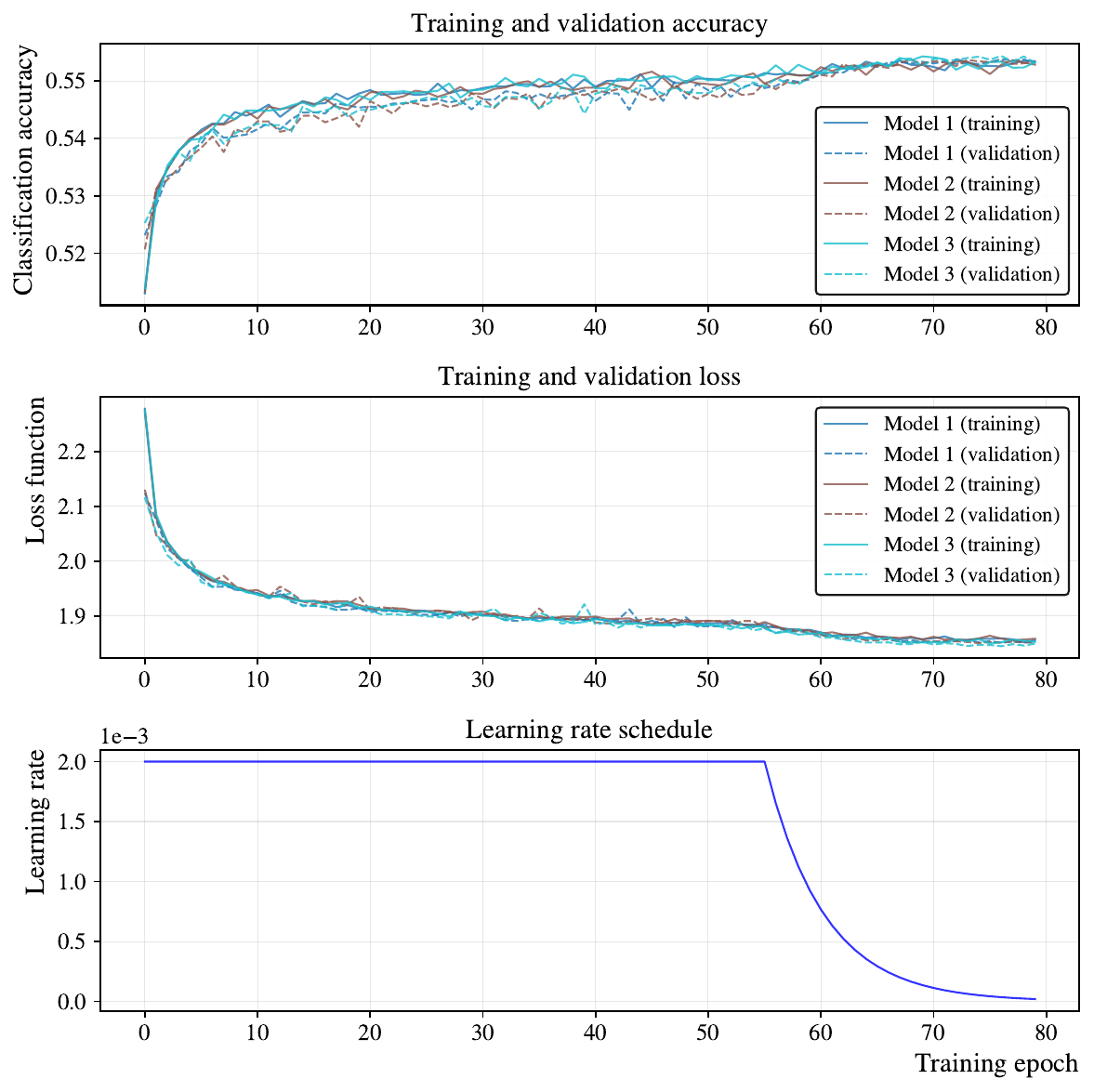}
\caption{Training performance metrics for the three-model ensemble. \textit{Top panel}: Classification accuracy evolution for training (solid lines) and validation (dashed lines) datasets. \textit{Middle panel}: Cross-entropy loss function convergence for training and validation samples. \textit{Bottom panel}: Learning rate schedule.}
\label{fig:training_metrics}
\end{figure}

\vspace{10pt}
\textbf{\textit{\boldmath Training performance.}}---
We find that all three ensemble members exhibit stable and consistent convergence over the 80 training epochs. Figure~\ref{fig:training_metrics} shows the evolution of key metrics, including classification accuracy, loss function, together with our learning rate scheduling.

As shown, the training and validation metrics remain closely aligned throughout the training process, indicating no evidence of overfitting despite the complexity of the 138-class classification task. All three ensemble members exhibit consistent convergence with minimal variation in their learning trajectories.

\subsection{Best-fit mass point determination}\label{app:fit-strategy}

Using the 136-dimensional output scores, we construct a discretized probability density $p(m_{h_1},\,m_{h_2})$. We then determine a best-fit mass prediction $(\hat{m}_{h_1},\,\hat{m}_{h_2})$, defined up to exchange symmetry, and assign the event to the corresponding grid cell. The purpose of this subsection is to describe the detailed implementation of this procedure.

It is first worth clarifying why a dedicated fitting strategy is considered helpful. If the probability density were unimodal, one could simply select the grid cell with the maximum probability as the final assignment, effectively performing a discretized ``regression''. However, the structure of $p(m_{h_1},\,m_{h_2})$ is more complex. By construction, it is symmetric with respect to the diagonal $m_{h_1}=m_{h_2}$. In cases where the distribution exhibits a double-peak structure symmetric about this diagonal, a naive maximum-probability criterion tends to merge the two peaks, making points along the diagonal artificially most probable. This leads to a systematic bias, manifesting as an overpopulation along the diagonal. Our goal is therefore to define a robust, rule-based method that avoids this pathology. Fitting the discrete $p(m_{h_1},\,m_{h_2})$ with a fixed analytical form provides a natural and effective solution.

We emphasize that the proposed fitting strategy offers additional advantages when it is applied to an experimental context. First, it maps the discrete probability density to a continuous mass estimate $(\hat{m}_{h_1},\,\hat{m}_{h_2})$, which provides greater flexibility for downstream analyses, such as rebinning of the mass histogram. Second, it mitigates the impact of statistical fluctuations in the neural network outputs. These outputs can be viewed as fluctuations around an underlying expectation, with larger variations arising from limited training data. A fitting-based estimator, rather than a simple maximum-probability selection, can reduce sensitivity to such fluctuations.

In the following, we present the details of the fitting strategy. While some aspects are technical, they can be further optimized in a full experimental implementation.

\vspace{10pt}
\textbf{\textit{\boldmath Event selection and discretized $p(m_{h_1},\,m_{h_2})$ reconstruction.}}---
Each event is first required to pass the CMS resolved ``4j3b'' trigger path, which demands four jets with $\pt > (75,\,60,\,45,\,40)$\GeV, $\HT > 330$\GeV, and at least three $b$-tagged jets. For events satisfying these criteria, the 136-dimensional classification output is converted into a discretized probability density $p(m_{h_1},\,m_{h_2})$, represented as a 256-bin histogram symmetric under $m_{h_1}\leftrightarrow m_{h_2}$. This is achieved by first populating the lower-triangular entries of a $16\times16$ matrix $H$ with the 136 scores, followed by symmetrization via the transformation $(H + H^{\mathrm{T}})/2$.

\vspace{10pt}
\textbf{\textit{\boldmath Two-stage double-sided crystal ball fitting strategy.}}---
The fitting procedure follows a two-stage strategy designed to identify potential multi-peak structures in the reconstructed probability density $p(m_{h_1},\,m_{h_2})$. The core of the method combines two-dimensional double-sided crystal ball (DCB) functions, which appear in symmetric pairs with respect to the diagonal $m_{h_1} = m_{h_2}$, together with a flexible background parameterization constructed to preserve the same diagonal symmetry.

A two-dimensional DCB function is defined as a direct extension of the standard one-dimensional DCB form through a rotational coupling,
\begin{equation}
\text{DCB}_{\text{2D}}(x,y) = A \cdot \text{DCB}_{\text{1D}}(x_{\text{rot}}) \cdot \text{DCB}_{\text{1D}}(y_{\text{rot}}),
\end{equation}
where $(x,y)$ denote the coordinates in the $(m_{h_1},\,m_{h_2})$ plane, and $(x_{\text{rot}}, y_{\text{rot}})$ are obtained by a rotation by an angle $\theta$. Each one-dimensional DCB component features asymmetric power-law tails characterized by the parameters $(\alpha_{\text{low}}, n_{\text{low}}, \alpha_{\text{high}}, n_{\text{high}})$. Since any peak structure describable by a DCB is necessarily symmetric under the exchange $x \leftrightarrow y$, the signal model is constructed as $\text{DCB}_{\text{2D}}(x,y) + \text{DCB}_{\text{2D}}(y,x)$.

The background contribution is modeled using a flexible functional form that combines linear and logarithmic terms,
\begin{equation}\label{eq:bkg-func}
B(x,y) = A_{\text{bg}} \left[1 + s_x(x-x_c) + s_y(y-y_c) + l_x\ln(x/x_c) + l_y\ln(y/y_c)\right],
\end{equation}
designed to capture the smooth, non-resonant behavior over the $[40,\,200]\GeV^2$ mass window. Diagonal symmetry is enforced by constructing the final background model as $B(x,y) + B(y,x)$.

The fitting is performed in two stages. In the first stage, a primary pair of DCB peaks, respecting exchange symmetry, is fitted together with the symmetric background model, with all parameters floated simultaneously to identify the dominant resonance structure. The optimization employs a robust curve-fitting procedure with bounded parameters: the DCB amplitudes are constrained to be positive, the mass positions are restricted to the physical range $[40,\,200]\GeV$, the width parameters are limited to $[2.5,\,50]\GeV$, and the tail parameters are subject to empirically determined bounds to ensure numerical stability.

After the primary peak is identified, residuals are computed by subtracting the fitted model from the original histogram. We then hope to identify residual peak structure from the remaining distribution. A masking procedure is first applied, excluding regions within $2\sigma$ of the primary peak and its symmetric counterpart. This allows for unbiased identification of secondary structures. Then, in the second stage, the same DCB-based fitting strategy is applied to the masked residual to extract secondary resonance structures, using the same parameter constraints.

\vspace{10pt}
\textbf{\textit{\boldmath Peak exchange and refinement logic.}}---
An important component of the method is a peak-exchange procedure designed to mitigate potential peak misidentification. If the primary peak is found outside the physical mass window $[40,\,200]\GeV$ while the secondary peak remains within bounds, the algorithm automatically exchanges the peak assignments and performs a localized refinement. In this refinement step, the geometric parameters of the newly promoted primary peak are fixed, while only its amplitude and the background parameters are re-optimized. This is followed by a full re-fitting of the secondary peak using the updated residual distribution. This exchange mechanism is essential for maintaining physical consistency, particularly in events where the initial fit converges to spurious local minima or where peak positions are shifted beyond the intended mass range due to imperfect detector resolution.

The primary peak position, denoted as $(\hat{m}_{h_1},\,\hat{m}_{h_2})$ and defined with the convention $\hat{m}_{h_1} \leq \hat{m}_{h_2}$, is taken as the mass prediction point obtained from our fitting strategy.

\begin{figure*}[ht!]
\centering
\includegraphics[width=\textwidth]{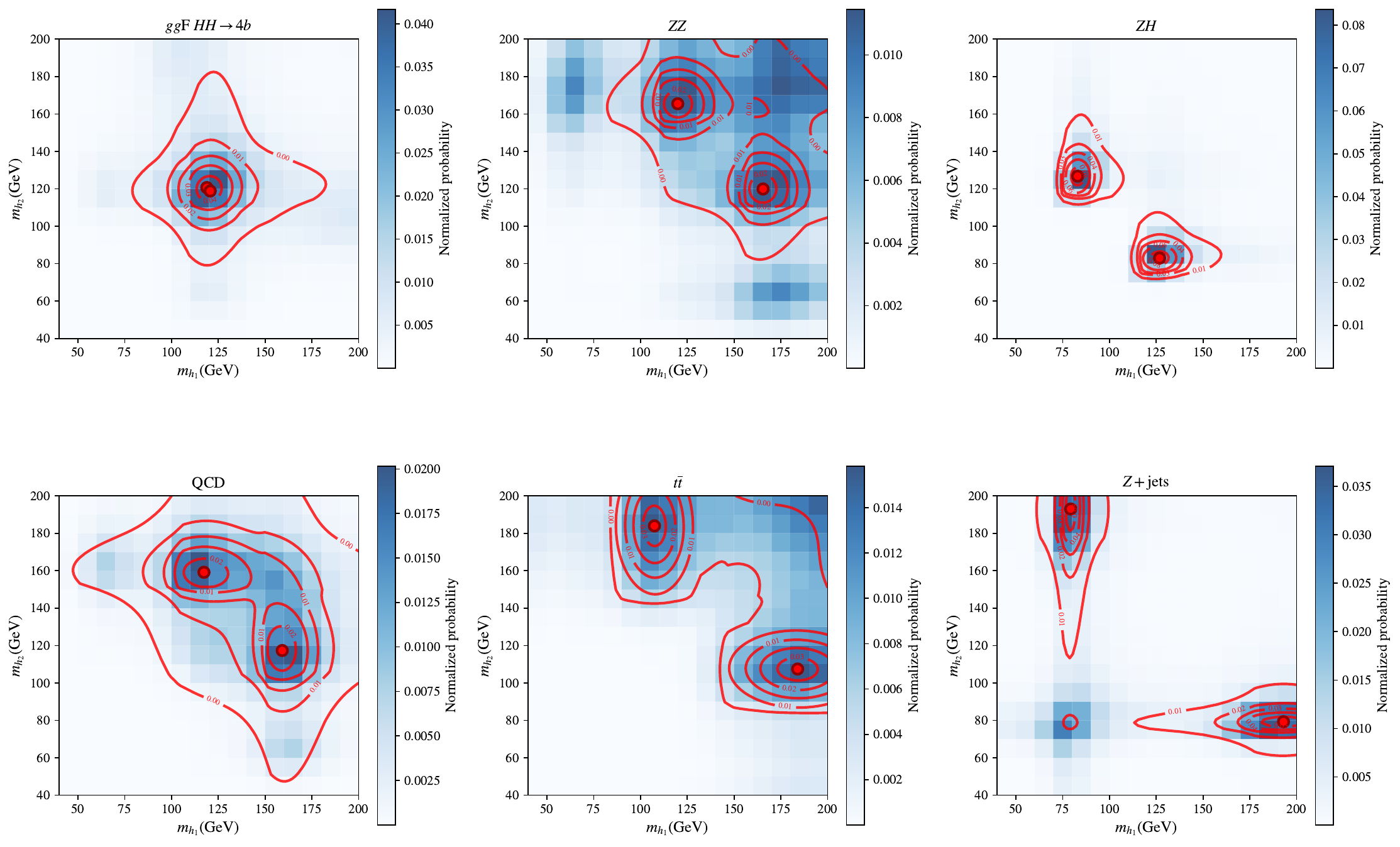}%
\caption{Single-event fitting validation showing $16\times16$ reconstructed $(m_{h_1},\,m_{h_2})$ histograms and corresponding fitted primary DCB functions (shown as contour lines) for representative events from major physics processes.}
\label{fig:fitting_validation}
\end{figure*}

\vspace{10pt}
\textbf{\textit{\boldmath Method validation.}}---
To demonstrate the reliability of the algorithm, Fig.~\ref{fig:fitting_validation} presents representative events from each of the six primary processes (\ggF $HH\to 4b$, $ZZ$, $ZH$, QCD multijet, \ttbar and $Z$+jets), together with their individual $16\times16$ two-dimensional mass histograms and the corresponding DCB functions obtained from the fit. The discretized $p(m_{h_1},\,m_{h_2})$ distributions are often nontrivial, exhibiting single-peak (considering one side of the diagonal) or more complex structures. Nevertheless, the proposed two-stage strategy successfully identifies genuine resonance features and provides a robust description of the underlying mass structure.

The peak position identified by our method is also indicated by the red dot in Fig.~\ref{fig:fitting_validation}. As can be seen, the fitting strategy successfully maps the discrete neural network outputs to continuous mass estimates, thus enabling the subsequent event-level analysis.

\subsection{Model ensembling strategy}
Each individual model provides a signal--background discrimination score $D_{h_1 h_2\to 4b}$, defined as the sum of the 136 signal-class output scores, as well as a mass prediction point $(\hat{m}_{h_1},\,\hat{m}_{h_2})$, denoted by $\hat{\mathbf{m}}$, obtained using the procedure described above and defined with the convention $\hat{m}_{h_1} \leq \hat{m}_{h_2}$. We combine the outputs of the three models using the following ensembling strategies to improve overall sensitivity and robustness:
\begin{itemize}
\item \textit{Discriminant}: The ensemble discriminant is defined as the arithmetic mean of the three model outputs,
\begin{equation}
D_{h_1 h_2\to 4b}^{\text{ens}} = \frac{1}{3} \sum_{i=1}^{3} D_{h_1 h_2\to 4b}^{(i)},
\end{equation}
where $i=1,2,3$ labels the individual models.
\item \textit{Mass prediction point}: The ensemble mass estimate is defined as the midpoint of the two closest prediction points,
\begin{equation}
\hat{\mathbf{m}}^{\text{ens}} = \frac{1}{2}(\hat{\mathbf{m}}^{(i^*)} + \hat{\mathbf{m}}^{(j^*)}), \qquad (i^*,j^*) = \arg\min_{i<j} \left| \hat{\mathbf{m}}^{(i)} - \hat{\mathbf{m}}^{(j)} \right|.
\end{equation}
\end{itemize}

Averaging the discriminant follows standard model-ensembling practice and consistently improves classification performance relative to any individual model (see Fig.~\ref{fig:ensemble_individual_roc}). The mass-prediction strategy is motivated by the observation that the fitting procedure may occasionally yield outlier solutions; selecting the two most consistent predictions substantially increases the likelihood of obtaining a reliable estimate. The ensemble approach improves both performance and robustness, and we believe it is well-suited for implementation in an experimental context.

\subsection{Mass window definition}\label{app:mass_window_def}

Based on the reconstructed ensemble mass prediction $\hat{\mathbf{m}}^{\text{ens}}=(\hat{m}_{h_1}^{\text{ens}}, \hat{m}_{h_2}^{\text{ens}})$, we populate histograms in the $m_{h_1}$--$m_{h_2}$ plane and define mass windows to evaluate the $HH$ sensitivity. When filling the histograms, diagonal symmetry is preserved by assigning half of each event to $(\hat{m}_{h_1}^{\text{ens}}, \hat{m}_{h_2}^{\text{ens}})$ and half to $(\hat{m}_{h_2}^{\text{ens}}, \hat{m}_{h_1}^{\text{ens}})$. As will be shown in Fig.~\ref{fig:mass_distribution_strategy_val}, although the best-fit strategy substantially reduces artificial population along the diagonal, a small residual enhancement remains due to imperfections in the fitting procedure. This residual structure invalidates the assumption of a smoothly varying QCD background.

To mitigate this effect, we apply an additional correction that uses a deterministic function to shift the mass prediction point along the direction perpendicular to the diagonal,
\begin{equation}
(\hat{m}_{h_1}^{\text{ens}},\, \hat{m}_{h_2}^{\text{ens}}) \;\to\; (\hat{m}_{h_1}^{\text{ens}} - \delta,\; \hat{m}_{h_2}^{\text{ens}} + \delta),
\end{equation}
where $\delta$ is a function of the average mass $M = (\hat{m}_{h_1}^{\text{ens}} + \hat{m}_{h_2}^{\text{ens}}) / 2$, mass difference $\Delta = |\hat{m}_{h_2}^{\text{ens}} - \hat{m}_{h_1}^{\text{ens}}|$ and the ``event tagger'' score $D_{h_1 h_2\to 4b}^{\text{ens}}$. The post-correction target distribution is constructed by summing the discretized joint distributions $p(m_{h_1},\, m_{h_2})$ output by the neural network on an event-by-event basis. More specifically, for each bin defined in $D_{h_1 h_2\to 4b}^{\text{ens}}$ and $M$, we consider the distribution of $\Delta$ and apply a cumulative distribution function (CDF) transformation to map it onto the corresponding post-correction target $\Delta$ distribution.
The resulting corrected distribution is illustrated in Fig.~\ref{fig:mass_distribution_strategy_val}~(d). In comparison with Fig.~\ref{fig:mass_distribution_strategy_val}~(c), the small diagonal over-density present before the correction is removed.



Figure~\ref{fig:mass_distributions} in the main text thus presents the distributions of the corrected mass prediction points for representative physics processes. As discussed there, the observed peak structures, ridge-like features, and smooth distributions in the $(m_{h_1},\,m_{h_2})$ plane are consistent with our expectations.

\begin{figure}[!ht]
\centering
\includegraphics[width=0.40\textwidth]{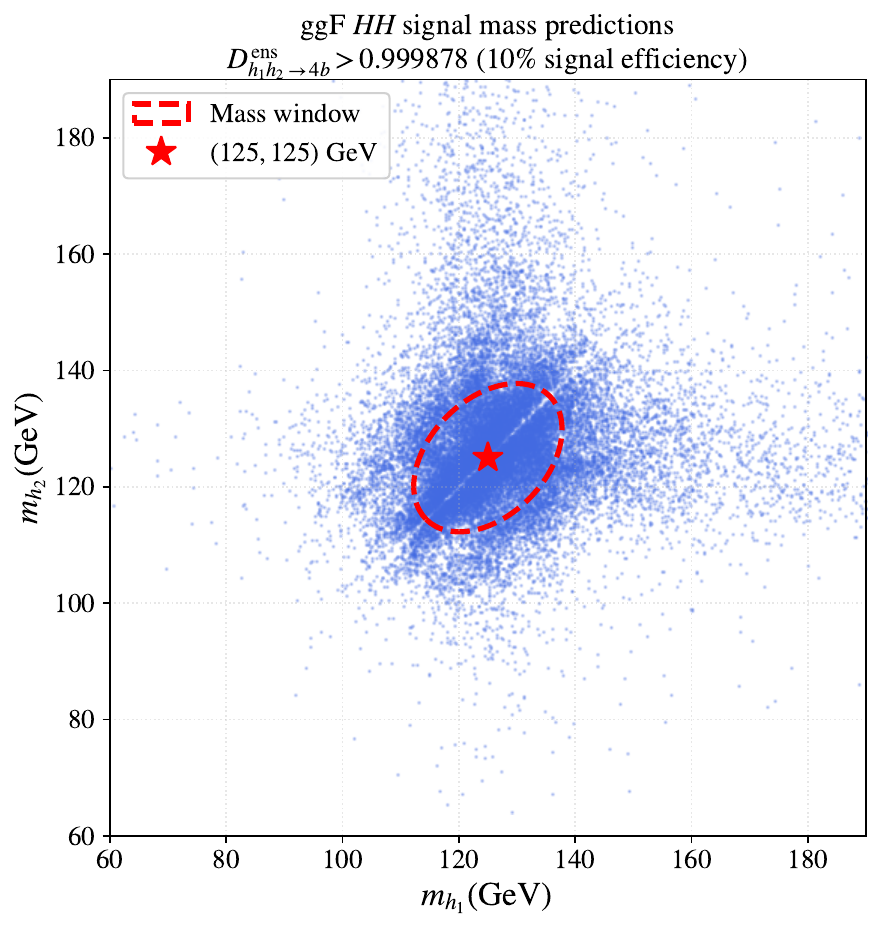}
\caption{Scatter plot of the corrected mass prediction points obtained with our strategy for the \ggF $HH$ signal, after applying a $D_{h_1 h_2\to 4b}^{\text{ens}}$ discriminant selection corresponding to a 10\% signal efficiency. The elliptical mass window used in the analysis is overlaid.}
\label{fig:signal_mass_window}
\end{figure}

To optimize the sensitivity, the mass-window selection is defined by an elliptical region centered at the Higgs mass point $(125,\,125)$\GeV, with semi-major axis $a = 15$\GeV aligned along the diagonal and semi-minor axis $b = 10$\GeV. For illustration, Fig.~\ref{fig:signal_mass_window} shows the distributions of the mass prediction points after applying the $D_{h_1 h_2\to 4b}^{\text{ens}}$ discriminant selections corresponding to \ggF signal efficiency of 10\%, with the elliptical mass window overlaid.

As a remark, we emphasize that our studies indicate the additional correction procedure, while \textit{ad hoc} in nature, does not affect the $HH$ search sensitivity, since the correction is applied uniformly to all processes. Even without applying the correction, corresponding to the QCD shape shown in Fig.~\ref{fig:mass_distribution_strategy_val}~(c), one can recover identical signal and background yields by a mild adjustment of the mass-window selection. The only purpose of the correction is therefore to smooth the QCD multijet distribution. We believe that more optimal mass-prediction strategies (potentially including better fit-based approaches with additional refinement) are possible; however, we defer such developments to future studies in the context of real experimental applications.

\subsection{Performance evaluation}\label{app:performance}

\subsubsection{Evaluation of \texorpdfstring{\ggF $HH$}{ggF HH} versus QCD separation performance}\label{app:hhvsqcd}

As illustrated by the red curve in Fig.~\ref{fig:roc_comparison}, we first study the discrimination power between the \ggF $HH$ signal and the QCD background using an ROC curve scaled to the total event yields. Events are required to pass the ``4j3b'' trigger, and the corrected mass prediction point must lie within the elliptical mass window. The event-tagger $D_{h_1 h_2 \to 4b}$ discriminant is then used to construct the ROC curve. Two technical aspects of the ROC-curve construction are addressed: (1) the enhancement of effective QCD background statistics by exploiting its approximately flat kinematic behavior in the $(m_{h_1},\,m_{h_2})$ plane, and (2) the impact of the model ensembling strategy.

On the first aspect, although the red curve is obtained after applying the elliptical mass-window requirement, at very tight $D_{h_1 h_2 \to 4b}$ working points (such as those used in our analysis), the effective number of QCD events surviving the selection falls below $\sim$5. This leads to large statistical fluctuations in the tail of the ROC curve and prevents a reliable estimate of the QCD background. To address this issue, we increase the effective background statistics by loosening the mass-window requirement for the QCD samples. Specifically, the ROC curve is constructed using the original signal sample restricted to the elliptical mass window, and QCD events selected with a looser circular mass window (centered at 125\GeV with a radius of 45\GeV). The resulting background yields are then rescaled to the elliptical mass-window region using a transfer function. The final background uncertainty, therefore, incorporates both the Poisson uncertainty associated with the enlarged background sample and the uncertainty in the fitted transfer factor.

We next assess the improvement achieved through model ensembling. For the ensemble case, the discriminant $D_{h_1 h_2 \to 4b}^{\text{ens}}$ is used as the event-tagger output, whereas the corrected $(\hat{m}_{h_1}^{\text{ens}},\, \hat{m}_{h_2}^{\text{ens}})$ values are employed to evaluate the mass-window requirement. This setup enables a direct comparison with the three individual models. We note that the mass-correction procedure is derived independently for each individual model as well as for the ensemble.

\begin{figure}[!ht]
\centering
\includegraphics[width=0.60\textwidth]{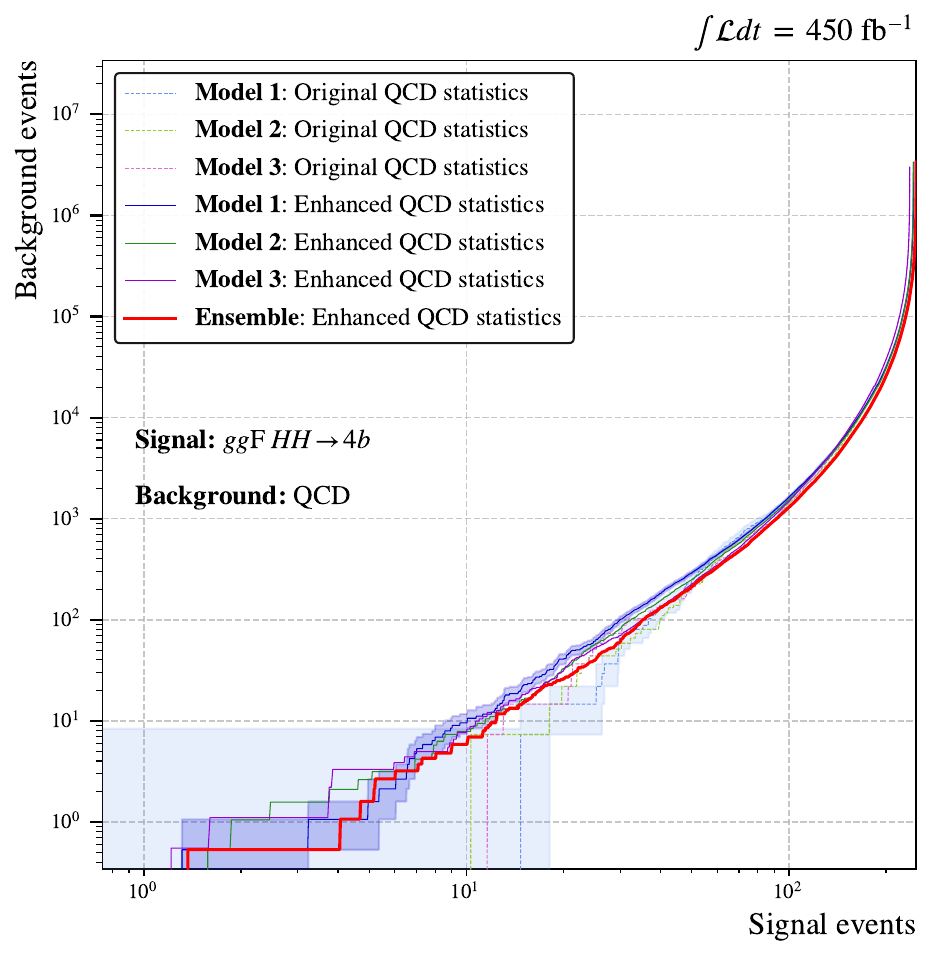}
\caption{ROC curves comparing the performance of individual ParT models and their ensemble combination for the \ggF $HH \to 4b$ signal against the QCD multijet background. The three individual models are shown in \textbf{blue, green, and purple}, while the ensemble is highlighted in \textbf{red}. \textbf{Solid lines (darker shades)} represent models trained on the full dataset with increased effective QCD statistics, whereas \textbf{dashed lines (lighter shades)} correspond to models trained on the original QCD samples. The dashed curves exhibit noticeable discrepancies in the extreme tail due to statistical limitations; however, these fluctuations remain consistent within the statistical uncertainty (bands shown only for Model 1/blue for clarity). With increased statistics, the curves become significantly more stable, and the ensemble discriminant consistently outperforms the best individual model across the entire phase space. The ensemble curve corresponds to the red line shown in Fig.~\ref{fig:roc_comparison} of the main text.}
\label{fig:ensemble_individual_roc}
\end{figure}

Figure~\ref{fig:ensemble_individual_roc} shows the performance of the three individual models and the ensemble, comparing the ROC curves obtained using the original QCD samples and those employing the increased effective QCD statistics. When using the original QCD events, the three models exhibit noticeable discrepancies in the extreme tail of the ROC curves, driven by statistical limitations. These differences are nevertheless consistent with the corresponding ROC uncertainties. After increasing the effective QCD statistics, the ROC curves become significantly more stable. Crucially, this comparison reveals that relying on the limited original statistics would result in a biased and overly optimistic estimation of the background rejection in the extreme tail. The ensemble discriminant consistently outperforms the best individual model across the entire signal--background phase space, demonstrating the effectiveness of the ensembling strategy in exploiting complementary model features. The ensemble ROC curve in Fig.~\ref{fig:ensemble_individual_roc} corresponds to the red curve shown in Fig.~\ref{fig:roc_comparison} of the main text.



\subsubsection{Characterization of the \texorpdfstring{$m_{h_1}$--$m_{h_2}$}{m_h1-m_h2} spectrum for QCD events}

We study how different strategies for deriving mass prediction points from the 136-dimensional network outputs affect the properties of the QCD $m_{h_1}$--$m_{h_2}$ spectrum. By construction, an optimal mass-point definition should produce a smooth QCD mass spectrum without inducing mass sculpting, since neither the signal training samples nor the QCD background samples are intrinsically localized around any particular mass value.

\begin{figure}[!ht]
\centering
\subfloat[]{\includegraphics[width=0.35\textwidth]{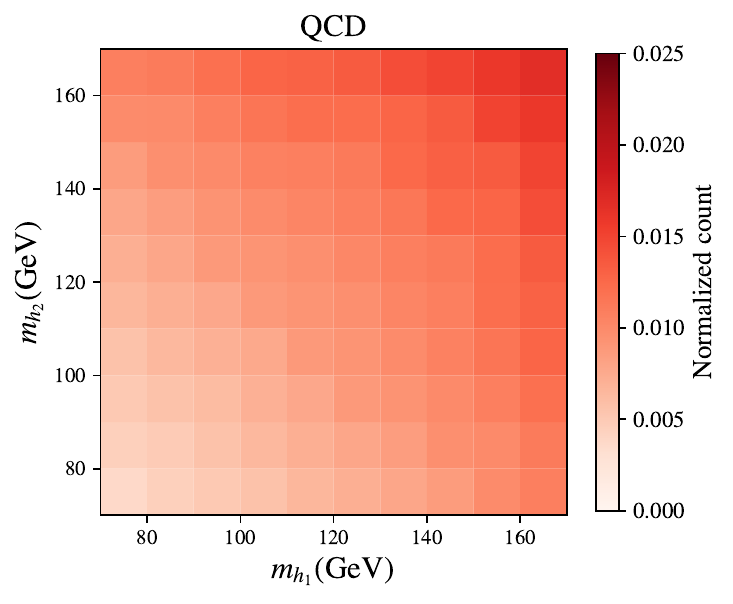}}
\subfloat[]{\includegraphics[width=0.35\textwidth]{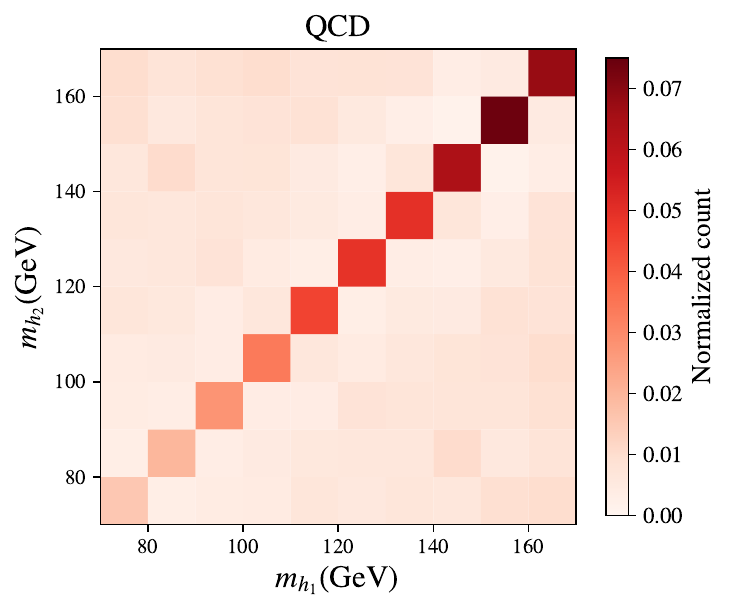}}\\
\vspace{-5pt}
\subfloat[]{\includegraphics[width=0.35\textwidth]{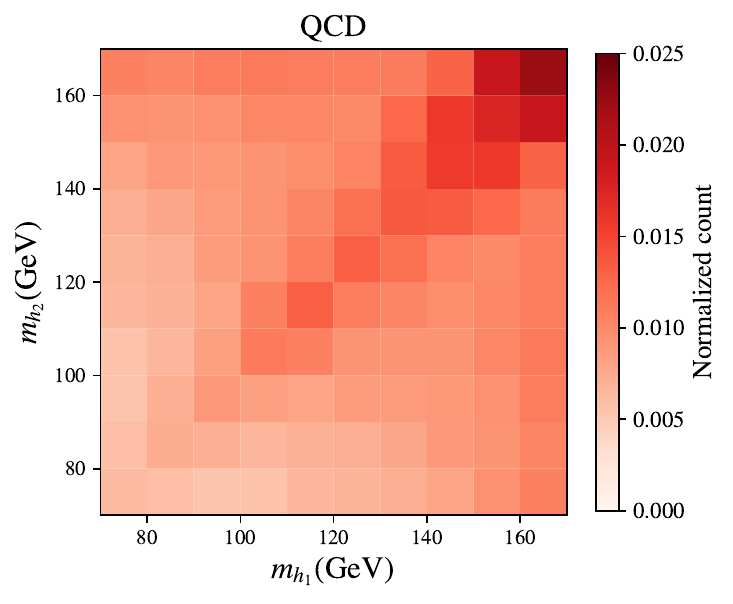}}
\subfloat[]{\includegraphics[width=0.35\textwidth]{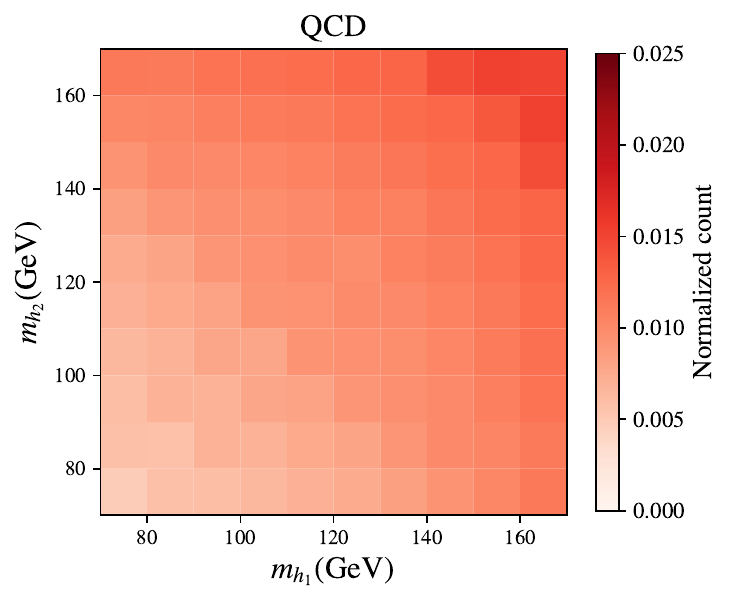}}
\caption{The $m_{h_1}$--$m_{h_2}$ spectra of QCD multijet events obtained using different mass-prediction strategies, with a basic selection $D_{h_1 h_2\to 4b} > 0.9$ applied. The histograms are constructed by (a) directly overlaying the learned discretized $p(m_{h_1},\,m_{h_2})$ scores, demonstrating the intrinsic smoothness of the QCD background; (b) filling each event at the bin corresponding to the maximum of $p(m_{h_1},\,m_{h_2})$; (c) filling the mass prediction points obtained from the fitting strategy in Appendix~\ref{app:fit-strategy}; and (d) filling the mass prediction points after applying the additional correction.}
\label{fig:mass_distribution_strategy_val}
\end{figure}

This expectation can be verified explicitly. When the 136-dimensional outputs are converted into 256-dimensional discretized $p(m_{h_1},\,m_{h_2})$ scores and directly summed over events, the resulting QCD background distribution is smooth, as shown in Fig.~\ref{fig:mass_distribution_strategy_val}~(a). Here and throughout this discussion, the QCD background is subject to a basic discriminant selection $D_{h_1 h_2\to 4b} > 0.9$, corresponding to the selection applied in Fig.~\ref{fig:mass_distributions}.

Next, we consider a naive mass-point assignment in which each event is filled in the $(m_{h_1},\,m_{h_2})$ bin corresponding to the maximum of $p(m_{h_1},\,m_{h_2})$. The resulting distribution, shown in Fig.~\ref{fig:mass_distribution_strategy_val}~(b), exhibits a pronounced accumulation along the diagonal. As discussed in Appendix~\ref{app:fit-strategy}, this effect arises because peak structures symmetric about the diagonal can merge into a single maximum, thereby biasing the selection toward diagonal points.

Figure~\ref{fig:mass_distribution_strategy_val}~(c) shows the $m_{h_1}$--$m_{h_2}$ distribution obtained using the more sophisticated fitting strategy described in Appendix~\ref{app:fit-strategy}. The diagonal enhancement is substantially reduced, although a small residual accumulation remains. Finally, Fig.~\ref{fig:mass_distribution_strategy_val}~(d) demonstrates the effect of the additional correction applied to further suppress the diagonal population introduced in Appendix~\ref{app:mass_window_def}, resulting in a smooth QCD mass spectrum.

This validation demonstrates the conceptual effectiveness of the fitting strategy introduced in Appendix~\ref{app:fit-strategy}, but it also reveals residual imperfections from the practical side. We believe further refinements can be pursued in realistic experimental contexts. Nevertheless, the present work provides a proof-of-principle framework for extracting the underlying probability densities $p(m_{h_1},\,m_{h_2})$ from multiclass classifiers and identifying the mass estimation points using a rule-based approach.

\subsubsection{Determination of optimal working points under full background modeling}\label{app:optimal-wp}

\begin{figure*}[ht!]
\centering
\includegraphics[width=0.60\textwidth]{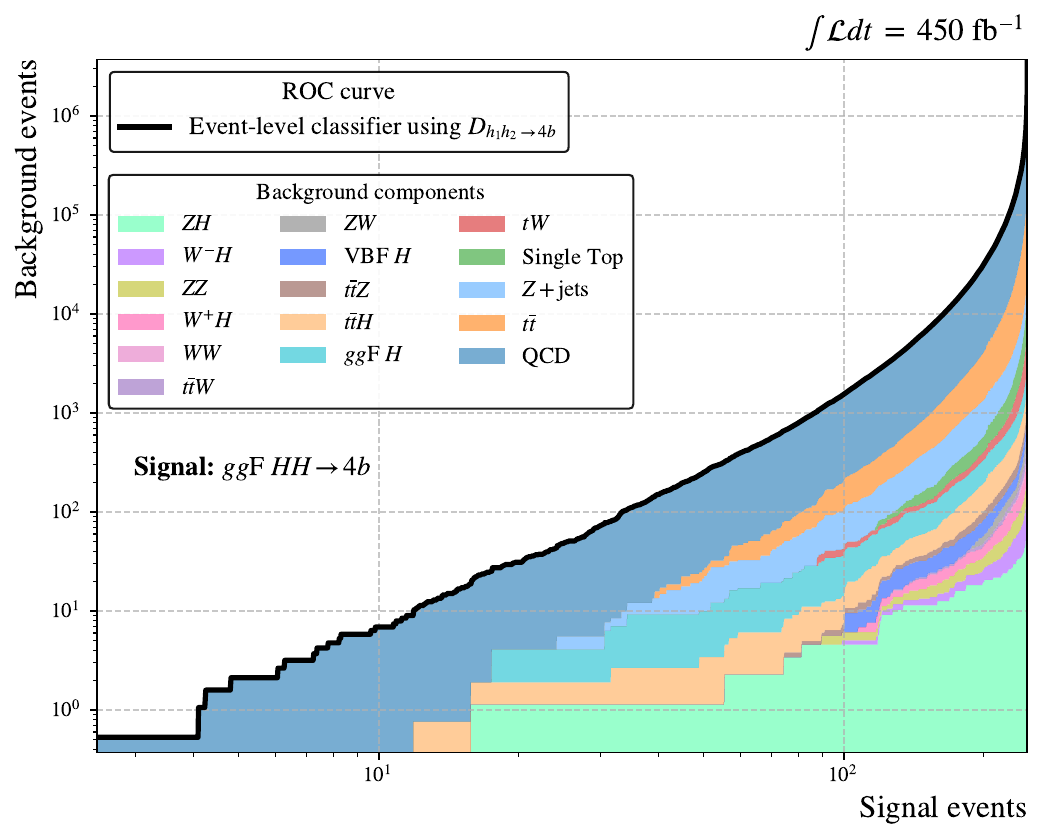}
\caption{Discrimination of the \ggF $HH$ signal from the total SM background as a function of the $D_{h_1 h_2\to 4b}^{\text{ens}}$ threshold, with individual background components shown as stacked colored bands. All event yields are normalized to an integrated luminosity of $450\invfb$. The optimal working point adopted in this analysis corresponds to expected signal and background yields of 20.5 and 32.7, respectively.}
\label{fig:roc_multi_process_std}
\end{figure*}

To fully assess the discriminative power of the proposed method, we perform a multi-process analysis that simultaneously tracks the signal efficiency and the composition of all SM background contributions. Using the complete background simulation summarized in Table~\ref{tab:sample_preparation}, Fig.~\ref{fig:roc_multi_process_std} shows the discrimination performance of the \ggF $HH$ signal against the total background, with individual background components displayed as stacked colored bands. The QCD background is still treated using the strategy described in Appendix~\ref{app:hhvsqcd} in order to enhance the effective statistics.

By optimizing the sensitivity, the selected working point corresponds to a threshold of $D_{h_1 h_2\to 4b}^{\text{ens}} = 0.9998996$. At this working point, the expected signal and background yields at an integrated luminosity of $450\invfb$ are 20.5 and 32.7, respectively. As shown, the QCD contribution remains dominant and exceeds the combined contributions of all other backgrounds by approximately an order of magnitude.

\subsubsection{Validation: impact of \texorpdfstring{\ttbar}{ttbar} background modeling in training}

A useful cross-check is to evaluate whether the \ttbar background must be explicitly vetoed as a dominant background component. This study is motivated by the observation that, in the reference CMS boosted analysis~\cite{CMS:2023yay}, \ttbar-induced large-$R$ jets are not explicitly vetoed, and the residual \ttbar contribution in the signal region remains sizable, even exceeding the QCD multijet background in some cases (see Fig.~1 of Ref.~\cite{CMS:2023yay}).

To investigate this effect, we define an alternative discriminant that does not explicitly suppress \ttbar contributions by removing them from the denominator,
\begin{align}
D_{h_1 h_2\to 4b}^{\mathrm{QCD}} = \frac{\sum_{i=1}^{136} p_i}{\sum_{i=1}^{136} p_i + p_{137}},
\end{align}
where $p_i$ denotes the network output for class $i$. For comparison, the nominal discriminant used in this analysis is
\begin{align}
D_{h_1 h_2\to 4b} = \frac{\sum_{i=1}^{136} p_i}{\sum_{i=1}^{136} p_i + p_{137} + p_{138}} = \sum_{i=1}^{136} p_i.
\end{align}
We note that defining $D_{h_1 h_2\to 4b}^{\mathrm{QCD}}$ is equivalent to performing a 137-class classification between the 136 signal classes and the QCD background only, with the sum of signal scores used as the discriminant.

\begin{figure*}[ht!]
\centering
\includegraphics[width=0.6\textwidth]{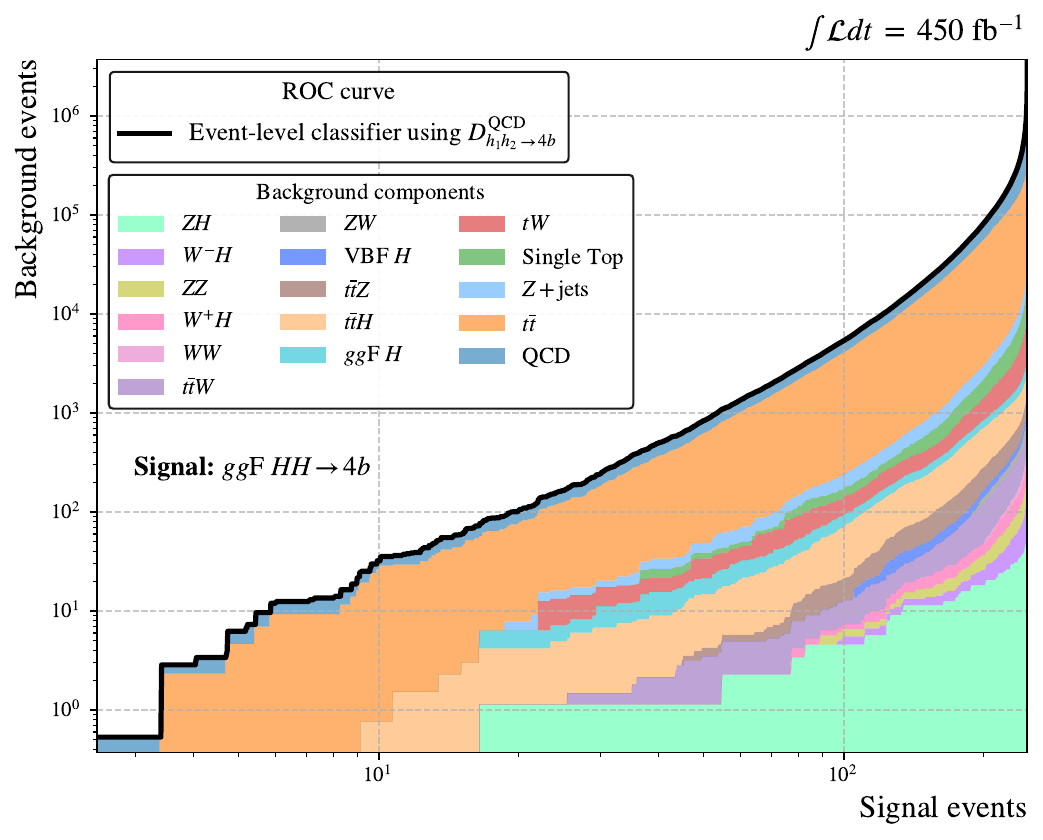}
\caption{Discrimination of the \ggF $HH$ signal from the total SM background as a function of the $D_{h_1 h_2\to 4b}^{\text{QCD,ens}}$ threshold, with individual background components shown as stacked colored bands. The discriminant $D_{h_1 h_2\to 4b}^{\text{QCD,ens}}$ is designed to suppress only the QCD multijet background, while leaving the \ttbar contribution unsuppressed.}
\label{fig:roc_multi_process_qcdonly}
\end{figure*}

Figure~\ref{fig:roc_multi_process_qcdonly} illustrates the resulting performance when $D_{h_1 h_2\to 4b}^{\mathrm{QCD}}$ is used. The background composition shows that, in the absence of an explicit \ttbar veto, the \ttbar contribution (orange bands) becomes substantial, in contrast to Fig.~\ref{fig:roc_multi_process_std}, where it is subdominant relative to QCD. At a representative signal efficiency corresponding to approximately 20 expected signal events, the \ttbar background yield is reduced from $\sim$\,74 to a negligible level when transitioning from $D_{h_1 h_2\to 4b}^{\mathrm{QCD}}$ to the nominal discriminant $D_{h_1 h_2\to 4b}$. In comparison, the QCD multijet contribution increases from 27 to 29, with no significant change.

This validation indicates that an explicit \ttbar veto can be critical when extending $HH \to 4b$ searches into more challenging regions of phase space. In particular, future boosted analyses, such as those in Refs.~\cite{CMS:2023yay,CMS-PAS-HIG-24-010}, may benefit from discriminants designed to suppress \ttbar-initiated jets explicitly.

\subsubsection{Validation: a baseline without sophisticated mass-point fitting}

We emphasize that the strong performance achieved in this work is driven by the intrinsic signal--background separation power of the discriminant $D_{h_1 h_2\to 4b}$, rather than by the use of a sophisticated mass-point fitting procedure. Even with the basic strategy, where each event is assigned to the bin with the maximum value of the discretized $p(m_{h_1},\,m_{h_2})$ (corresponding to Fig.~\ref{fig:mass_distribution_strategy_val}~(b)), one can already identify working points that are very sensitive to $HH\to 4b$. In this case, however, all background components exhibit clear accumulation along the diagonal, making background estimation challenging.

\begin{figure*}[ht!]
\centering
\includegraphics[width=0.6\textwidth]{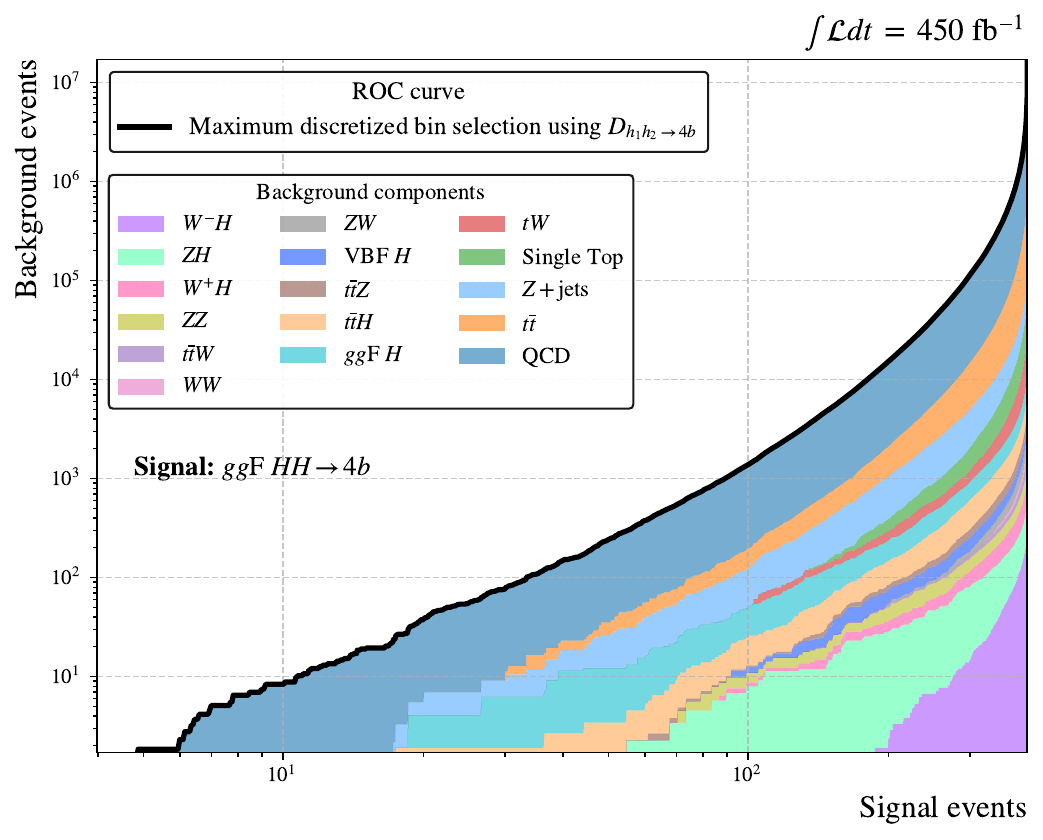}
\caption{Discrimination of the \ggF $HH$ signal from the total SM background as a function of the $D_{h_1 h_2\to 4b}^{\text{ens}}$ threshold, using a simplified mass-prediction strategy that assigns each event to the bin with the maximum discretized $p(m_{h_1},\,m_{h_2})$. Three diagonal bins centered on the Higgs mass are selected. Interestingly, the resulting performance is already comparable to that shown in Fig.~\ref{fig:roc_multi_process_std}.}
\label{fig:roc_multi_process_maxbin}
\end{figure*}

Figure~\ref{fig:roc_multi_process_maxbin} illustrates the performance obtained with this simple mass-prediction strategy. The mass region is selected using a minimal window defined by three diagonal bins around the Higgs boson mass,
\begin{equation}
    [110,\,120]\times [110,\,120],  \quad [120,\,130]\times [120,\,130], \quad  [130,\,140]\times [130,\,140]\;\GeV^2
\end{equation}

The ensemble discriminant $D_{h_1 h_2\to 4b}^{\text{ens}}$ is still used to separate the signal from all SM backgrounds. A similar approach that relies on an enlarged mass window to enhance the effective QCD statistics is adopted. As a result, at a \ggF $HH$ signal yield of 20 events, the expected background reaches 37.5, which is already close to the optimal working points discussed in Appendix~\ref{app:optimal-wp}.

This study demonstrates that the discriminant alone provides substantial sensitivity. Nevertheless, a more sophisticated mass-prediction strategy enables continuous $(\hat{m}_{h_1},\,\hat{m}_{h_2})$ estimates, which provide more flexibility in optimizing the mass window shape, and thereby further improving the $HH$ sensitivity.

\subsubsection{Validation: ablation studies with reduced training data}\label{app:val-reduced-data}

In the main text, we emphasized that the key ingredient supporting our jet-free model's performance is the availability of an extensive training dataset. As an ablation study, we examine the impact of reducing the training statistics by a factor of 10.

In this study, all signal and background samples are simultaneously downscaled by a factor of 10. The model is trained with the same architecture and hyperparameter configuration as in the nominal setup, with only the training schedule modified. Specifically, the number of mini-batches per epoch is reduced by a factor of 2.5 (from $2.56\times10^{6}$ to $1.02\times10^{6}$), and the total number of epochs is reduced by a factor of 2 (from 80 to 40). Thus, the total number of sample iterations is reduced by a factor of 5. This configuration is optimized to allow the model to undergo as many training iterations as possible before overfitting occurs. The resulting performance of the reduced-statistics model is shown by the orange curve in Fig.~\ref{fig:roc_comparison}. Clearly, its performance is substantially inferior to that of the nominal model (red curve), with the discrepancy increasing in an even tighter selection regime: at the working point corresponding to 20 expected signal events, the nominal model achieves a QCD background suppression that is a factor of $\sim$\,5 stronger than that of the reduced-statistics model. At a more stringent working point corresponding to 10 signal events, this advantage further increases to a factor of $\sim$\,7.8. These observations demonstrate that \textit{the size of the training dataset plays a decisive role in determining the model performance}.

This behavior can be well-understood. In the resolved channel, the QCD background must be suppressed by about seven orders of magnitude, from an initial yield of $3.5\times10^{8}$ events to $\mathcal{O}(30)$ events after all selections. The model must therefore learn to identify the extremely rare $\mathcal{O}(10^{-7})$ region of the QCD phase space that most closely resembles the genuine $HH\to4b$ signal, while rejecting the remaining $1-10^{-7}$ fraction. Achieving such discrimination requires sufficiently large QCD statistics precisely in this extreme tail of phase space in order to obtain an adequate description of the low-level feature distributions. This consideration implies that the original QCD event sample, after the resolved trigger selection, must contain at least $\mathcal{O}(10^{7}$\text{--}$10^{8})$ events. In addition, our further study indicates that the scaling behavior has not yet saturated. We therefore anticipate that the achievable performance, even when based on our high-fidelity fast-simulation samples, may exceed the results reported here.

It is worth emphasizing that the $\mathcal{O}(10^{8})$ training samples used in this work are already comparable to those employed in current ATLAS and CMS analyses for training central, highly complex Transformer-based jet-tagging models~\cite{ATLAS:2025dkv,CMS-DP-2024-066}. Moreover, since this dataset is used to develop an analysis-level background suppression tool, it requires generating a much larger number of unfiltered events (e.g., $\mathcal{O}(10^{11})$ raw QCD events in our case). Producing datasets of this scale using full detector simulation within the ATLAS or CMS workflows is clearly challenging. For future experimental deployment of this approach, it may be advantageous to pursue strategies that increase the effective training statistics, e.g., by combining fast- and full-simulation samples during training to achieve effective sample sizes of $\mathcal{O}(10^{8})$ or larger.

\subsubsection{Validation: \texorpdfstring{\ggF $HH$}{ggF HH} versus QCD separation across various Lorentz boosts}

\begin{figure*}[!ht]
\centering
\includegraphics[width=0.50\textwidth]{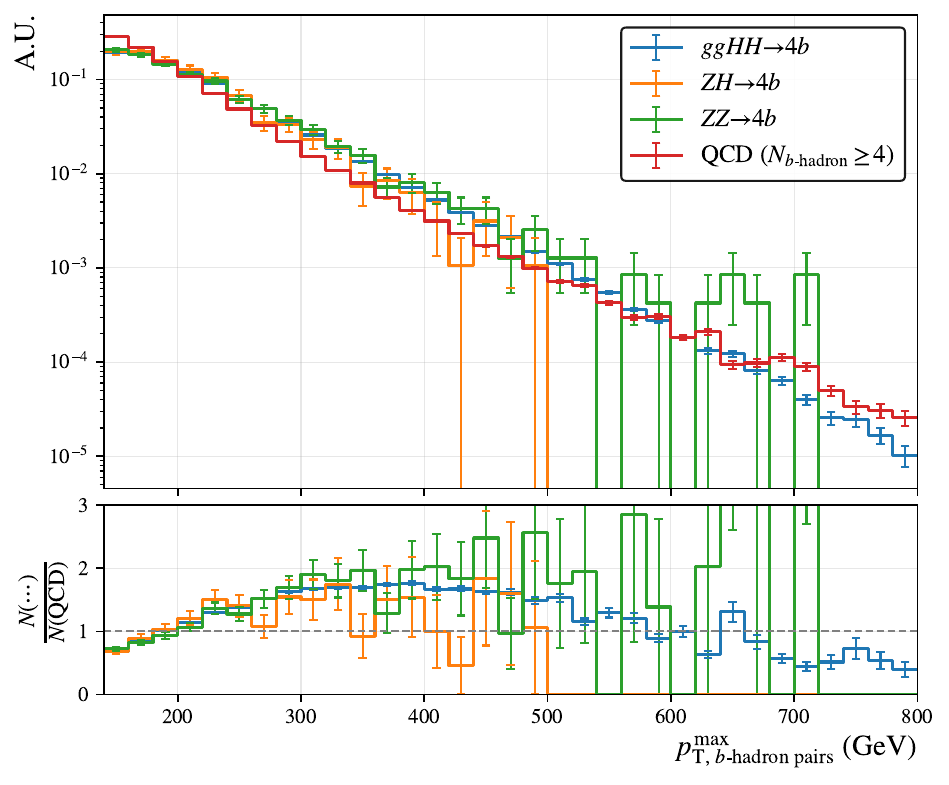}
\caption{Distribution of the maximum \pt among all possible pairings of the four $b$ hadrons ($p_{\text{T,\,$b$-hadron pairs}}^{\max}$), used as a proxy for the event Lorentz boost, for three processes: \ggF $HH$, $ZZ$, and QCD multijet processes with $\geq 4$ $b$ hadrons. All distributions are normalized to unity for events satisfying $p_{\text{T,\,$b$-hadron pairs}}^{\max} > 140\GeV$. Error bars represent statistical uncertainties.}
\label{fig:validation_lorentz_boost}
\end{figure*}

We aim to validate whether the discriminant $D_{h_1 h_2\to 4b}$ exhibits comparable separation power between \ggF $HH$ signal and QCD background across different Lorentz-boost regimes. In both the ATLAS and CMS experiments, it is traditionally expected that searches in the boosted regime achieve enhanced signal--background separation, since signal processes tend to populate higher \pt regions (often featuring longer $\pt$ tail) while backgrounds do not. As a result, the signal-to-background ratio is naturally improved at high $\pt$. In the context of $HH$ production, this argument holds for searches targeting deviations in $\kappa_{2V}$, where modifications to the Higgs self-interaction significantly alter the $m_{HH}$ spectrum and drive the Higgs bosons to higher $\pt$, making the boosted regime particularly sensitive~\cite{CMS:2023yay}. However, for SM \ggF $HH$ production, this expectation is not apparent, as its $\pt$ spectrum is not substantially ``harder'' than that of the background. This behavior is illustrated in Fig.~\ref{fig:validation_lorentz_boost}.

Here, to quantify the event boost in a way that is uniformly applicable to both signal and background, we restrict the QCD multijet sample to events containing exactly four $b$ hadrons, allowing a direct comparison with the $HH \to 4b$ final state. We define the Lorentz boost proxy as the maximum \pt among all possible pairings of the four $b$ hadrons,
$p_{\text{T,\,$b$-hadron pairs}}^{\max}$. Using this observable, Fig.~\ref{fig:validation_lorentz_boost} compares the distributions for \ggF $HH$, $ZH$, and $ZZ \to 4b$, and QCD multijet processes. Due to the resolved-trigger requirements, only events with $p_{\text{T,\,b-had pairs}}^{\max} > 140\GeV$ are considered. After normalizing the event yields across the three samples, we observe that the characteristic $\pt$ spectra of $HH$, $ZH$, and $ZZ$ do not exhibit harder tails compared to the QCD background.

These observations suggest that the strong $HH$ versus QCD separation power achieved in the boosted regime by CMS can be effectively extended to the broader resolved phase space. To demonstrate this, we evaluate the performance of the $D_{h_1 h_2\to 4b}$ discriminant in different Lorentz-boost intervals, which is shown in Fig.~\ref{fig:validation} (left) in the main text. We also compare this performance to that obtained by replacing $D_{h_1 h_2\to 4b}$ with the boosted-region event discriminant based on \Sophon, defined in Eq.~(\ref{eq:sophon-discr}), in the highest-$\pt$ interval. Interestingly, the two approaches yield comparable separation power, which confirms that the performance achieved in the boosted regime can be effectively transferred to the full phase space.

\subsubsection{Validation: impact of hadronization modeling and \texorpdfstring{$ZZ\to 4b$}{ZZ->4b} spin correlations}

To assess the robustness of the proposed jet-free signal--background discrimination model against theoretical modeling uncertainties, we perform validation studies that examine the impact of hadronization modeling and spin correlations on the classifier discriminant distributions.

\vspace{10pt}
\textbf{\textit{\boldmath Hadronization modeling.}}---
Following standard validation procedures for jet-tagging studies at the LHC~\cite{CMS:2020poo,ATLAS:2025dkv}, we evaluate the stability of the event-level discriminant using three parton-shower generators: \PYTHIA~8.3~\cite{Sjostrand:2014zea}, \HERWIG~7.2~\cite{Bellm:2019zci}, and \VINCIA~\cite{Fischer:2016vfv}. For this study, the \ggF $HH$ signal samples are generated starting from identical hard-scattering events produced with \POWHEG, with the different hadronization models applied subsequently. The total cross sections are normalized to the same value for all generators.

\begin{figure*}[!ht]
\centering
\includegraphics[width=0.42\textwidth]{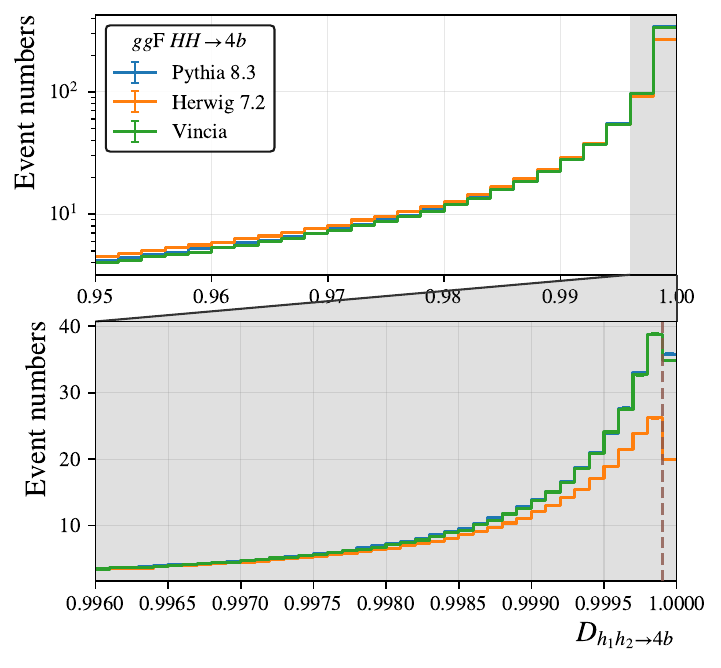}\\
\includegraphics[width=0.42\textwidth]{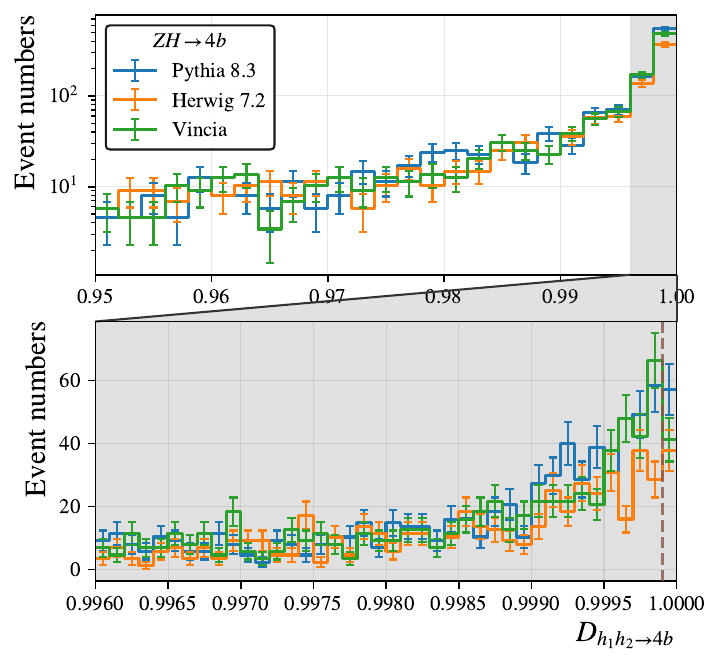}\hspace{10pt}
\includegraphics[width=0.42\textwidth]{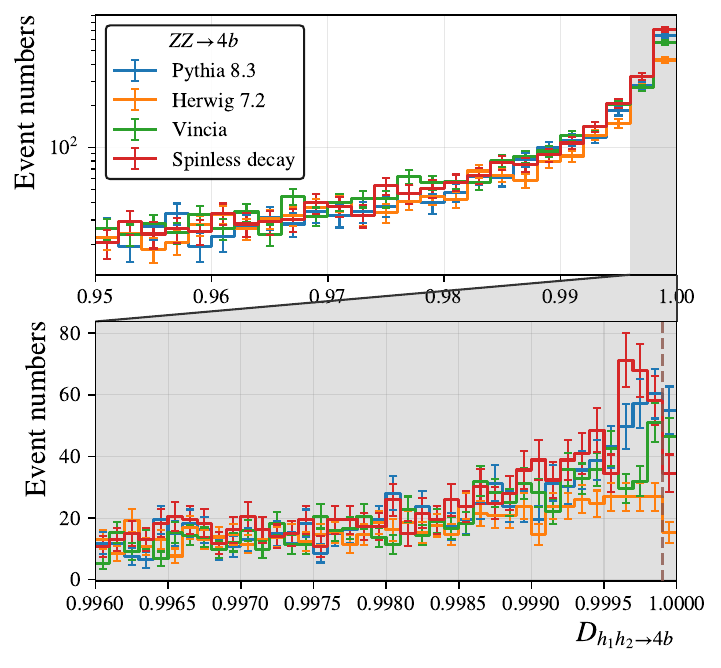}
\caption{\textit{Top:} $D_{h_1 h_2 \to 4b}$ discriminant distributions for \ggF $HH \to 4b$ events across three parton shower generators. \textit{Bottom left:} Distributions for $ZH \to 4b$ events. \textit{Bottom right:} Distributions for $ZZ \to 4b$ events, including comparison with spinless decay (red) to validate calibration transfer. The vertical line in the lower panels indicates the $D_{h_1 h_2 \to 4b}$ threshold defining the chosen working point. All results correspond to 450\invfb integrated luminosity. Error bars represent statistical uncertainties.}
\label{fig:generator_discriminant_validation}
\end{figure*}

Figure~\ref{fig:generator_discriminant_validation} shows the $D_{h_1 h_2 \to 4b}$ distributions for the \ggF $HH \to 4b$ (top), $ZH \to 4b$ (bottom left), and $ZZ \to 4b$ (bottom right) processes obtained with the three generators. In the high-discriminant region, i.e., when the score exceeds 0.996, finer binning is employed to enhance sensitivity to potential modeling differences under very tight selection requirements. The $y$-axes correspond to the expected yields for the respective processes for an integrated luminosity of $450~\invfb$.

As observed, the \PYTHIA and \VINCIA models exhibit similar behavior, while larger differences are seen with respect to \HERWIG. This pattern is consistent with experimental observations, where \HERWIG-based hadronization is known to differ noticeably from \PYTHIA in the performance of traditional large-$R$ jet taggers~\cite{ATL-PHYS-PUB-2023-020}. Such differences are therefore expected. This study supports the conclusion that the proposed jet-free approach, like large-$R$ jet taggers already used in experiments, can be reliably employed in data analyses. In the experimental context, calibrations are derived by applying data-to-simulation scale factors based on \PYTHIA-hadronized samples, which typically provide a closer description of the data (see, e.g., Refs.~\cite{CMS:2025kje,CMS-DP-2025-010}). Since our calibration strategy relies exclusively on \PYTHIA-based simulations through the $ZZ$ process and does not depend on \HERWIG samples, the observed \PYTHIA--\HERWIG differences do not introduce additional systematic uncertainties.

\vspace{10pt}
\textbf{\textit{\boldmath Spin correlation modeling.}}---
To validate the transfer of the calibration from $ZZ \to 4b$ to $HH \to 4b$, we study the impact of boson spin on the classifier response. Figure~\ref{fig:generator_discriminant_validation} (bottom right) additionally compares standard $ZZ \to 4b$ events, involving spin-1 $Z$ bosons, with hypothetical spin-0 $ZZ$ decays generated with identical kinematics but scalar couplings (red histogram). The latter sample is produced by generating the hard process $pp \to ZZ$ with \MGvATNLO and subsequently decaying the $Z$ bosons to $b\overline{b}$ using \PYTHIA, which enforces a spin-0 decay hypothesis for the $Z \to b\overline{b}$ decay.

The resulting discriminant distributions exhibit negligible sensitivity to spin correlations among the final-state $b$ quarks. This validates the use of $ZZ \to 4b$ as a calibration proxy for the $HH \to 4b$ signal and confirms that differences in angular correlations among the four $b$ quarks do not have a significant impact on the discrimination performance.

\section{Supplementary details on sensitivity estimation and \texorpdfstring{$\kappa_\lambda$}{kappa_lambda} constraints}\label{app:signal-extraction}

Based on the extracted signal and background yields presented in Appendix~\ref{app:optimal-wp}, we estimate the sensitivity using the Asimov approximation and extract constraints on $\kappa_\lambda$ via a profiled likelihood-ratio fit. It is necessary to quantify the corresponding uncertainties on the signal and background estimates, denoted by $\sigma_s$ and $\sigma_b$.

\subsection{Determination of uncertainty schemes}

We adopt a fit-based strategy to explore a realistic scenario on signal and background uncertainties. In our proposed methodology, the $HH\to 4b$ signal can be calibrated using the $ZZ\to 4b$ process. As a result, a simultaneous fit to the $ZZ$ resonance peak and the smooth QCD multijet background allows the extraction of the respective uncertainties on the signal and background yields. This treatment is particularly well motivated, as QCD multijet production remains the dominant background under the chosen working points.

\begin{figure*}[!ht]
\centering
\includegraphics[width=0.96\textwidth]{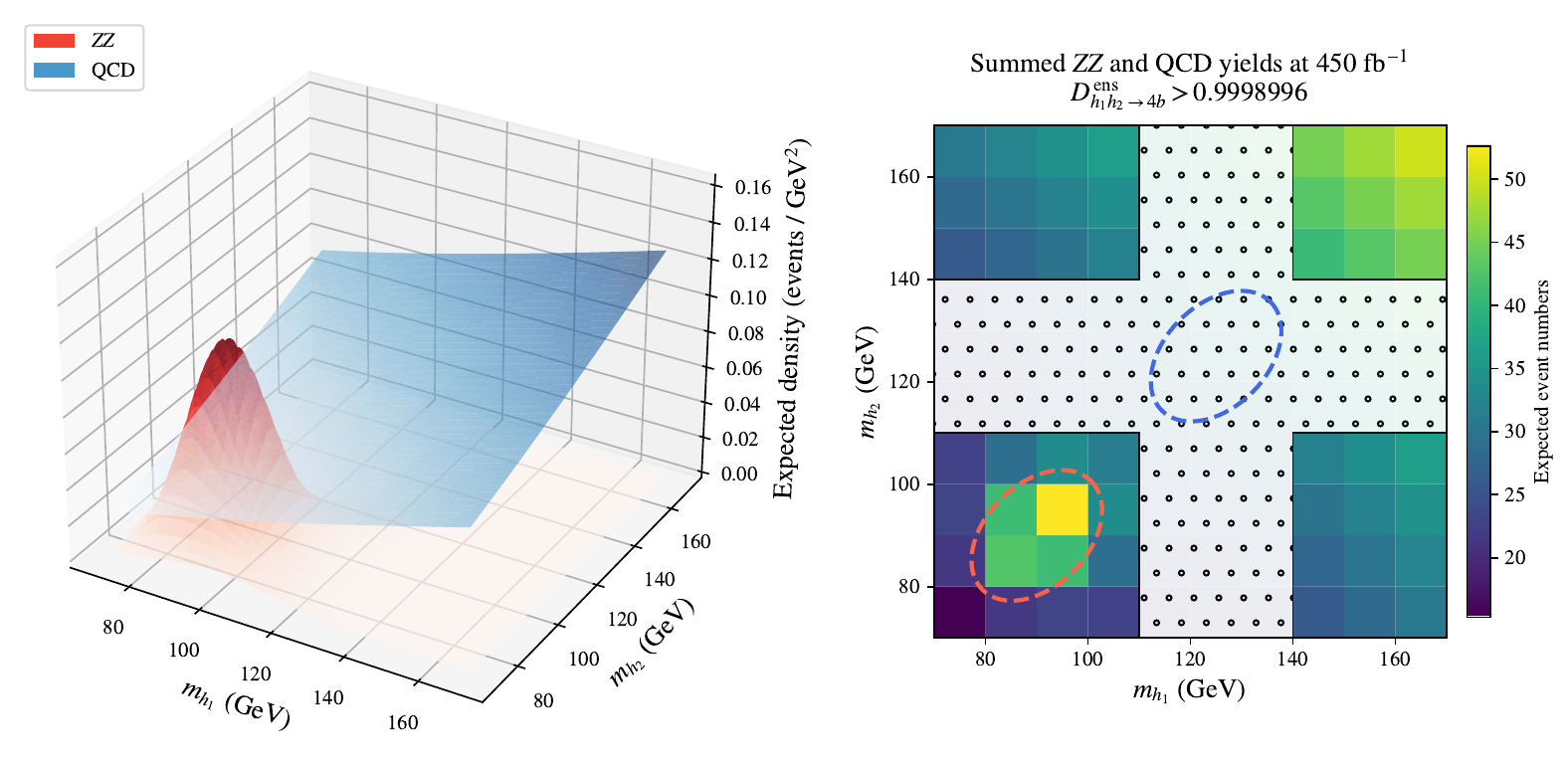}
\caption{\textit{Left:} Pre-extracted probability density functions for the $ZZ$ signal and the QCD background on the $(m_{h_1},\,m_{h_2})$ plane in the $[70,170]\times[70,170]\;\mathrm{GeV}^2$ range, used in the fit-based study to estimate signal and background uncertainties. Each component is normalized to its expected yield after applying the $D_{h_1 h_2 \to 4b}$ selection defining the working point.
\textit{Right:} Expected yields of the $ZZ$ and QCD components in each bin of the binned $(m_{h_1},\,m_{h_2})$ distribution. Regions excluded from the fit are masked with dots in the histogram. The elliptical mass windows used to estimate the $ZZ$ signal yield in the $ZZ$ vicinity and the QCD background yield in the $HH$ vicinity are indicated. Event yields correspond to an integrated luminosity of 450\invfb.}
\label{fig:unc_scheme_estimate}
\end{figure*}

The fit is performed in a restricted region of the $(m_{h_1},\,m_{h_2})$ plane, retaining only the vicinity of the $(m_Z,\,m_Z)$ resonance while explicitly excluding the $(m_Z,\,m_H)$, $(m_H,\,m_Z)$, and $(m_H,\,m_H)$ regions in order to eliminate contamination from $ZH$ and $HH$ processes. This fit region is illustrated in Fig.~\ref{fig:unc_scheme_estimate} (right), with the masked regions indicated by dots. For the $ZZ$ signal and the QCD multijet background, we assume that their shapes in the $(m_{h_1},\,m_{h_2})$ distribution follow predefined probability density functions. The $ZZ$ contribution is modeled by a two-dimensional DCB function, centered along the diagonal, with the two axes defined parallel and perpendicular to the diagonal direction. The QCD background is described by a simple smooth function, as given in Eq.~(\ref{eq:bkg-func}). The total yields of these components in the $[70,170]\times[70,170]\;\mathrm{GeV}^2$ region are taken from simulation, corresponding to approximately 64.0 $ZZ$ events and 738 QCD events. The corresponding two-dimensional probability density functions are shown in Fig.~\ref{fig:unc_scheme_estimate} (left), where the vertical axis represents the event density per $\mathrm{GeV}^2$, while the resulting distributions on the discrete $(m_{h_1},\,m_{h_2})$ grid are displayed in Fig.~\ref{fig:unc_scheme_estimate} (right).

The signal and background uncertainties are extracted as follows. Using the joint probability density function described above, toy datasets are generated with the total number of events drawn from a Poisson distribution centered on the expected yields. An unbinned likelihood fit is then performed on each toy dataset, keeping the shapes of the $ZZ$ and QCD components fixed while allowing only their yields to float. This procedure yields best-fit values for the $ZZ$ and QCD normalizations. We subsequently record the best-fit yields of the $ZZ$ signal and the QCD background in mass windows around the $ZZ$ peak and the $HH$ signal region, respectively, as illustrated in Fig.~\ref{fig:unc_scheme_estimate} (right). By repeating this procedure across many toy experiments, we obtain the standard deviations of these yields, which are approximately 18.8\% for the $ZZ$ component and 6.4\% for the QCD background.

These results serve as guidance for the uncertainty scheme adopted in the subsequent analysis. As a realistic choice, we assume a 20\% uncertainty in the signal yield (with $ZZ$ serving as a proxy for the $HH$ signal) and a 10\% uncertainty in the QCD background yield.

It is worth noting that the quoted $ZZ$ and QCD uncertainties are tied to the current discriminating power of the $D_{h_1 h_2 \to 4b}$ discriminant. Improved separation between $ZZ$ and QCD would allow for an optimal working point with higher $ZZ$ acceptance or stronger QCD rejection, thereby further reducing the uncertainties on the $ZZ$ and QCD yields extracted with this method. In a realistic experimental implementation, the actual performance of $D_{h_1 h_2 \to 4b}$ is expected to be even stronger than illustrated in this study. This expectation is motivated, first, by the fact that the \Sophon $bb$-tagger employed here has inferior performance compared to the CMS \texttt{ParticleNet-MD} or \texttt{GloParT} taggers (as shown in Table~\ref{tab:bkgrej_sophon}); second, the strong scaling behavior of the discriminant performance with training data size indicates that the results presented here have not yet reached saturation (as discussed in Appendix~\ref{app:val-reduced-data}). Thus, the achievable uncertainties in real data are expected to be smaller than those estimated in this section.

\subsection{Sensitivity estimation}

Based on the considerations above, we define four uncertainty schemes corresponding to different assumptions on the relative uncertainties of the signal and background yields in the signal region:
\begin{itemize}
    \item \textit{statistical-only}, with no systematic uncertainties;
    \item \textit{optimistic}, assuming a 5\% background uncertainty together with a 10\% signal uncertainty;
    \item \textit{realistic}, assuming a 10\% background uncertainty together with a 20\% signal uncertainty;
    \item \textit{conservative}, assuming a 20\% background uncertainty together with a 30\% signal uncertainty.
\end{itemize}
The $HH$ search sensitivity is evaluated using the Asimov approximation,
\begin{equation}
Z_A = \sqrt{2 \left\{ (s + b) \ln \left[\frac{(s + b)(b + \sigma_b^2)}{b^2 + (s + b)\sigma_b^2}\right] - \frac{b^2}{\sigma_b^2} \ln \left[1 + \frac{\sigma_b^2 s}{b(b + \sigma_b^2)}\right] \right\}},
\end{equation}
where $s$ and $b$ denote the expected signal and background yields, respectively, and $\sigma_b$ is the background uncertainty. The resulting sensitivities for the different uncertainty scenarios are summarized in Table~\ref{tab:results} in the main text.

\subsection{Likelihood scan in \texorpdfstring{$\kappa_\lambda$}{kappa_lambda} and constraint extraction}

Constraints on the Higgs self-coupling modifier $\kappa_\lambda$ are derived using a profile-likelihood approach implemented within a multidimensional fitting framework.

\vspace{10pt}
\textbf{\textit{\boldmath Signal process modeling.}}---
An accurate modeling of the $HH$ production cross-section as a function of $\kappa_\lambda$ is a key ingredient of the likelihood analysis. The cross-section is parameterized as a linear combination of matrix elements corresponding to different coupling configurations. For \ggF $HH\to 4b$, the cross-section follows a three-component decomposition:
\begin{equation}
\sigma_{\ggF}(\kappa_\lambda, \kappa_t) = A_\text{box} \kappa_t^4 + A_\text{tri} \kappa_t^2 \kappa_\lambda^2 + A_\text{int} \kappa_t^3 \kappa_\lambda
\end{equation}
where $A_\text{box}$, $A_\text{tri}$, and $A_\text{int}$ represent the box diagram, triangle diagram, and interference contributions, respectively.

For VBF $HH\to 4b$, a six-component parameterization captures the dependence on $C_V$, $C_{2V}$, and $\kappa_\lambda$:
\begin{equation}
\sigma_{\mathrm{VBF}}(C_V, C_{2V}, \kappa_\lambda) = \sum_{i=1}^{6} B_i \cdot f_i(C_V, C_{2V}, \kappa_\lambda)
\end{equation}
where the functions $f_i$ include terms such as $C_V^2\kappa_\lambda^2$, $C_V^4$, $C_{2V}^2$, $C_V^3\kappa_\lambda$, $C_V C_{2V} \kappa_\lambda$, and $C_V^2 C_{2V}$.

The matrix element coefficients are determined by solving a linear system constructed from Monte Carlo samples generated at specific coupling points. For \ggF $HH\to 4b$ production, three benchmark points suffice to determine the three coefficients, while VBF $HH\to 4b$ requires six benchmark configurations.

\vspace{10pt}
\textbf{\textit{\boldmath Event yield configuration.}}---
The signal and background event yields used in the likelihood construction are summarized in Table~\ref{tab:event-yields}. These yields correspond to the optimal discriminant threshold identified in Appendix~\ref{app:optimal-wp} and assume 450\invfb integrated luminosity.

\begin{table}[htb]
\centering
\caption{Event yields for different di-Higgs production processes and background contributions used in the $\kappa_\lambda$ constraint derivation. All yields correspond to the optimal score threshold and 450~fb$^{-1}$ integrated luminosity.}
\label{tab:event-yields}
\begin{tabular}{lcc}
\toprule
\textbf{Process} & \textbf{Coupling configuration} & \textbf{Event yield} \\
\midrule
\multicolumn{3}{c}{\textit{Gluon-gluon fusion}} \\
\ggF $HH\to 4b$ & $\kappa_\lambda = 1$, $\kappa_t = 1$ & 20.511 \\
\ggF $HH\to 4b$ & $\kappa_\lambda = 0$, $\kappa_t = 1$ & 29.230 \\
\ggF $HH\to 4b$ & $\kappa_\lambda = 5$, $\kappa_t = 1$ & 8.663 \\
\midrule
\multicolumn{3}{c}{\textit{Vector boson fusion}} \\
VBF $HH\to 4b$ & $C_V = 1$, $C_{2V} = 1$, $\kappa_\lambda = 1$ & 0.118 \\
VBF $HH\to 4b$ & $C_V = 1$, $C_{2V} = 2$, $\kappa_\lambda = 1$ & 39.847 \\
VBF $HH\to 4b$ & $C_V = 1$, $C_{2V} = 1$, $\kappa_\lambda = 2$ & 0.067 \\
VBF $HH\to 4b$ & $C_V = 1$, $C_{2V} = 1$, $\kappa_\lambda = 0$ & 0.485 \\
VBF $HH\to 4b$ & $C_V = 0.5$, $C_{2V} = 1$, $\kappa_\lambda = 1$ & 24.376 \\
VBF $HH\to 4b$ & $C_V = 1.5$, $C_{2V} = 1$, $\kappa_\lambda = 1$ & 81.080 \\
\midrule
\multicolumn{3}{c}{\textit{Background}} \\
Total background & -- & 32.654 \\
\bottomrule
\end{tabular}
\end{table}

\vspace{10pt}
\textbf{\textit{\boldmath Likelihood scan configuration.}}---
The profile likelihood scan over $\kappa_\lambda$ employs a grid-based approach with 100 evaluation points spanning the range $\kappa_\lambda \in [-5,\,15]$. At each grid point, the likelihood is maximized with respect to all nuisance parameters while keeping $\kappa_\lambda$ fixed. The other coupling parameters ($\kappa_t$, $C_V$, $C_{2V}$) are held at their SM values.

The statistical model incorporates systematic uncertainties through log-normal nuisance parameters affecting both signal and background normalizations. We consider four uncertainty scenarios, where background (signal) uncertainties are set as $0\,(0)\%$, $5\,(10)\%$, $10\,(20)\%$, and $20\,(30)\%$, following the discussion above. For each scenario, the profile likelihood ratio $\Lambda(\kappa_\lambda) = -2\ln[\mathcal{L}(\kappa_\lambda)/\mathcal{L}(\hat{\kappa}_\lambda)]$ is computed, where $\hat{\kappa}_\lambda$ denotes the best-fit value. The 95\% confidence intervals are determined by the condition $\Lambda(\kappa_\lambda) = 3.84$, corresponding to the critical value for one degree of freedom.

\begin{figure}[htb]
\centering
\includegraphics[width=0.55\textwidth]{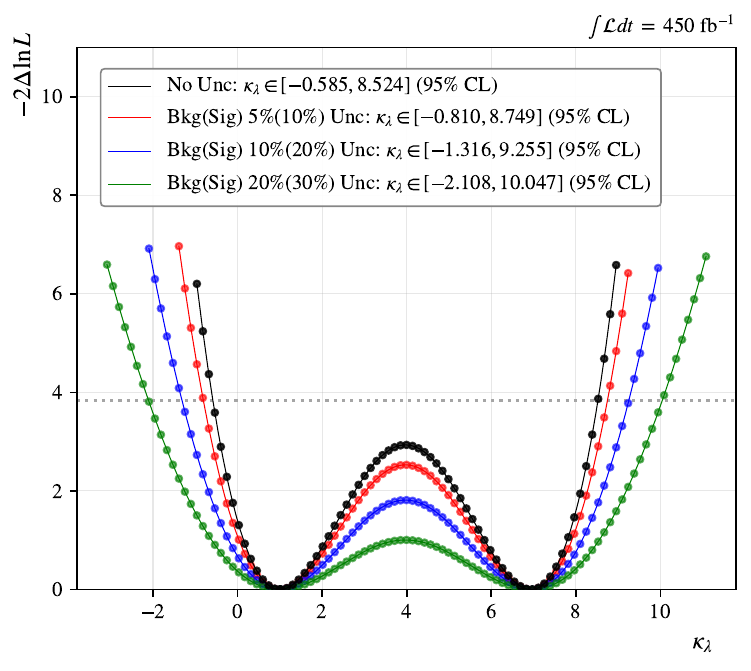}
\caption{Profile likelihood scans for $\kappa_\lambda$ under four systematic uncertainty scenarios. The horizontal dashed line indicates the 95\% confidence level threshold ($-2\Delta\ln\mathcal{L} = 3.84$). The double-minimum structure arises from destructive interference effects in the $\mathrm{\ggF}~HH\to 4b$ production mechanism.}
\label{fig:kappa-lambda-scan}
\end{figure}

\vspace{10pt}
\textbf{\textit{\boldmath Results.}}---
Figure~\ref{fig:kappa-lambda-scan} presents the profile likelihood scans for all four systematic uncertainty scenarios. The corresponding 95\% confidence interval is reported in Table~\ref {tab:results} in the main text. Here, the characteristic double-minimum structure arises from the destructive interference between box and triangle contributions in \ggF $HH\to 4b$ production. This leads to two distinct regions of parameter space consistent with the observed event yield. Overall, these constraints represent a significant improvement over current experimental sensitivity.

\end{document}